\documentclass[10pt,twocolumn,letterpaper]{article}

\usepackage{cvpr}              
\usepackage{graphicx}
\usepackage{tikz}
\usepackage{float} 
\usepackage{pdfpages}
\usepackage{comment}
\usepackage{wrapfig}
\usepackage{amsmath}
\usepackage[ruled]{algorithm2e} %
\usepackage{lipsum} 
\usepackage{tcolorbox} 
\usepackage{listings}
\usepackage{xcolor}
\usepackage{shadowtext}

\usepackage{wrapfig}

\definecolor{cvprblue}{rgb}{0.21,0.49,0.74}
\usepackage[pagebackref,breaklinks,colorlinks,allcolors=cvprblue]{hyperref}

\def\confName{CVPR}
\def\confYear{2025}

\title{BlenderGym: Benchmarking Foundational Model Systems for Graphics Editing }

\author{Yunqi Gu\\
{\normalsize Stanford University} \\
{\tt\small yrichard@stanford.edu}
\and
Ian Huang\\
{\normalsize Stanford University} \\
{\tt\small ianhuang@stanford.edu}
\and
Jihyeon Je\\
{\normalsize Stanford University} \\
{\tt\small jihyeonj@stanford.edu}
\and
Guandao Yang\\
{\normalsize Stanford University} \\
{\tt\small guandao@stanford.edu}
\and
Leonidas Guibas\\
{\normalsize Stanford University}\\
{\tt\small guibas@cs.stanford.edu}
}

\begin{document}
\twocolumn[{%
\renewcommand\twocolumn[1][]{#1}%
\maketitle
\begin{center}
\includegraphics[width=0.9\linewidth, trim=230 160 240 35, clip]{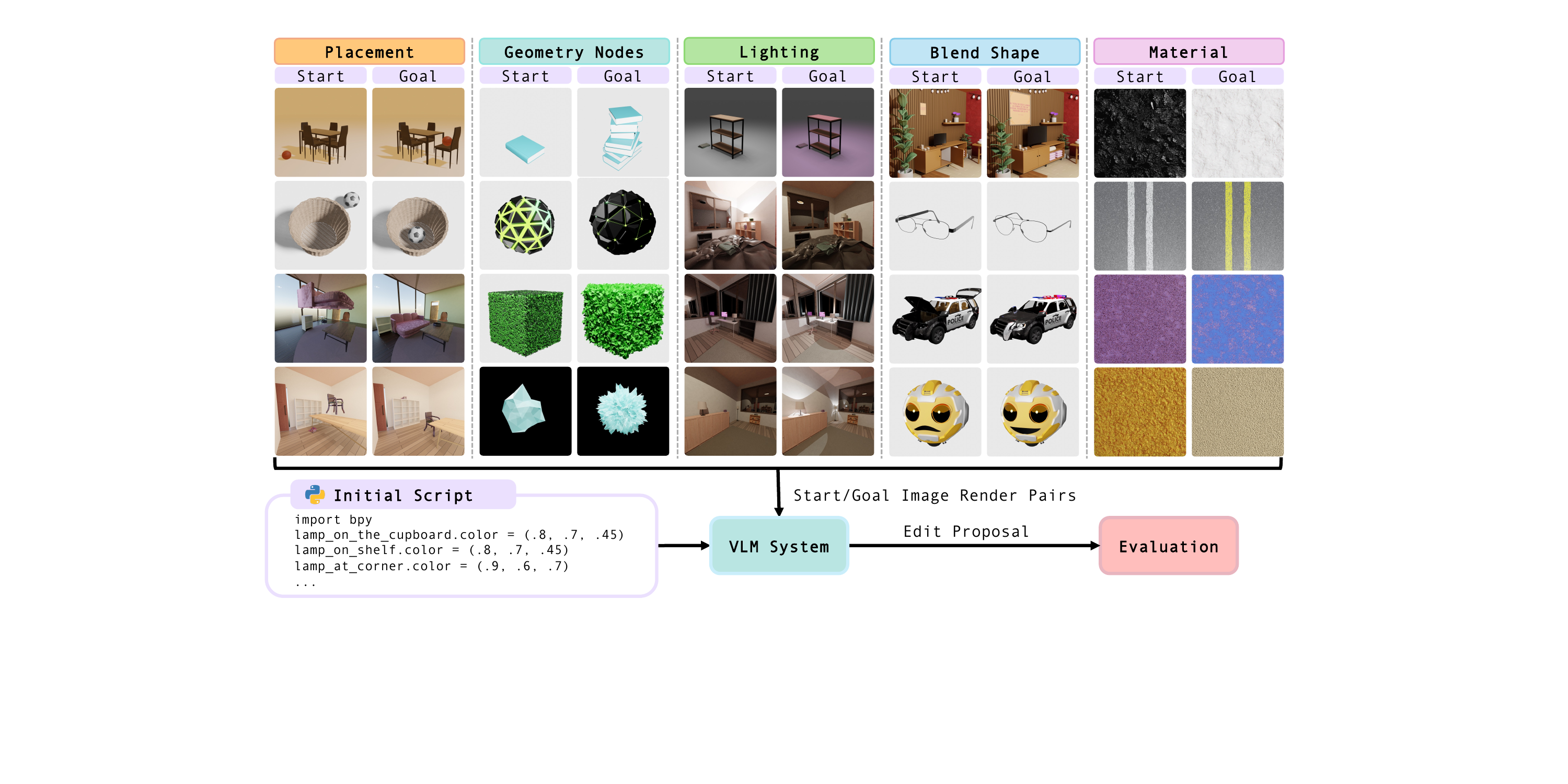}
\end{center}
\vspace{-2em}
\captionof{figure}{Example task instances of BlenderGym, a 3D graphics benchmark that tasks VLMs with 3D scene reconstruction through code editing. BlenderGym consists of 245 handcrafted start-goal scene pairs across five key graphics editing tasks: object placement, lighting adjustment, procedural material editing, blend shape manipulation, and procedural geometry editing.}
\label{fig: teaser}
\vspace{1em}
}]

\begin{abstract}
3D graphics editing is crucial in applications like movie production and game design, yet it remains a time-consuming process that demands highly specialized domain expertise. 
Automating this process is challenging because graphical editing requires performing a variety of tasks, each requiring distinct skill sets. 
Recently, vision-language models (VLMs) have emerged as a powerful framework for automating the editing process, but their development and evaluation are bottlenecked by the lack of a comprehensive benchmark that requires human-level perception and presents real-world editing complexity. 
In this work, we present \textbf{BlenderGym}, the first comprehensive VLM system benchmark for 3D graphics editing. BlenderGym evaluates VLM systems through code-based 3D reconstruction tasks. We evaluate closed- and open-source VLM systems and observe that even the state-of-the-art VLM system struggles with tasks relatively easy for human Blender users. 
Enabled by BlenderGym, we study how inference scaling techniques impact VLM's performance on graphics editing tasks. 
Notably, our findings reveal that the verifier used to guide the scaling of generation can itself be improved through inference scaling, complementing recent insights on inference scaling of LLM generation in coding and math tasks. We further show that inference compute is not uniformly effective and can be optimized by strategically distributing it between generation and verification. 
\end{abstract}

\section{Introduction}\label{sec:intro}

Producing 3D graphics for video games and movies is tedious and challenging. Beyond mastering 3D software such as Blender, Unity, Unreal, and Maya, artists must be skilled in diverse tasks, each with its own workflow—from animation to material design and geometry modeling. To address these diverse and difficult tasks, significant efforts have focused on automating graphics editing to accelerate production and reduce barriers to entry.

Prior works \cite{de2024llmr, sun20233dgpt, yamada2024lgo, goel2024iterative, yang2024holodeck, feng2024layoutgpt, huang2023aladdin, hu2024scenecraft, fu2024anyhome} demonstrate that large language models \cite{team2023gemini, jiang2023mistral, 2023GPT4} can automate tasks in 3D graphics editing, such as layout generation, animation sequence creation, and texture generation. Building on this, \citet{huang2024blenderalchemy} explore using vision-language models (VLMs) \cite{2023GPT4VisionSC} to edit 3D procedural geometry, material, and lighting by leveraging their visual understanding to guide edits. As multi-modal foundation models continue advancing, they could potentially play an increasingly central role in 3D graphics editing. In contrast to the growing applications of VLMs in the field, their evaluation remains underexplored, making it difficult for the community to compare different approaches and identify where improvements are needed.



Despite the growing demand for evaluation metrics for graphics, prior attempts\cite{huang2024blenderalchemy, yang2024holodeck, yamada2024lgo, de2024llmr, sun20233dgpt, hu2024scenecraft} are not comprehensive enough. Some primarily rely on limited sample sets within the narrow task settings in which their method specializes and, therefore, do not allow cross-comparison between different approaches~\cite{huang2024blenderalchemy, hu2024scenecraft, sun20233dgpt}. Others are limited by the lack of reliable, cost-effective, and quantitative metrics~\cite{hu2024scenecraft, huang2024blenderalchemy, yamada2024lgo, yang2024holodeck, chen2024mllm, lee2024prometheus, li2024leveraging, wu2024meta, chen2024mj}. Unlike fields such as mathematics, graphics editing generally lacks standardized answers, making human evaluation a common choice~\cite{hu2024scenecraft, huang2024blenderalchemy, yamada2024lgo, yang2024holodeck}, which is often slow and expensive. Some works, therefore, have introduced the AI-as-judge paradigm~\cite{chen2024mllm, lee2024prometheus, li2024leveraging, wu2024meta, chen2024mj}, but this raises concerns about reliability due to models' lack of reasoning capabilities and, more critically, the bias when the same AI judge model is also involved in the pipeline being judged.

To this end, we introduce BlenderGym (\cref{fig: teaser}), a 3D graphics editing benchmark for VLM systems. BlenderGym evaluates VLM systems by tasking them with reconstructing a goal Blender scene from a start Blender scene through Python program editing. It (1) covers a wide variety of task settings in 3D graphics editing, (2) allows easy plug-and-play of customized VLM systems, and (3) provides fixed start-goal scene pairs to enable quantitative evaluation of editing results, removing human or AI involvement in evaluation.
BlenderGym covers five key task settings of 3D graphics editing: object placement, lighting adjustment, procedural material editing, blend shape manipulation, and procedural geometry editing. These tasks are chosen to meet the demand of downstream graphics editing applications and can therefore reflect a VLM system's abilities in the field. We also provide human baselines to contextualize the performance of VLM systems and demonstrate the solvability and validity of each task. 


In addition to VLM system evaluation, BlenderGym enables us to study inference scaling to enhance VLM systems' graphics editing capabilities. 
As shown in \cite{brown2024largelanguagemonkeysscaling, snell2025scaling}, scaling inference compute can be effective \textit{if} a capable \textit{verifier} exists to select among a large set of candidate answers reliably. 
Specifically, given that VLMs can function as verifiers that select among their own outputs based on visual comparisons~\cite{huang2024blenderalchemy}, we demonstrate on BlenderGym that 
(1) VLMs used as verifiers to guide generation could also benefit from inference scaling, complementing their established scaling laws as generators~\cite{brown2024largelanguagemonkeysscaling}, and that (2) inference compute is not uniformly effective and can be optimized by strategically allocating it between VLM generation and verification. When more compute is available, a higher share of verification yields greater benefits.


In summary, our contributions are as follows:
\begin{itemize}
\item  We introduce BlenderGym, a VLM system benchmark on 3D graphics that tasks VLMs with code-based 3D scene reconstruction. It evaluates VLM systems quantitatively on various realistic editing tasks, spanning across lighting, material, blend shape, geometry, and placement. 
\item We provide an evaluation of state-of-the-art (SOTA) closed- and open-source VLM systems on BlenderGym, as well as a comparison to human baselines.
\item We apply our benchmark to study inference time scaling for graphics editing, particularly the effectiveness of scaling the verifier in the paradigm introduced in \cite{brown2024largelanguagemonkeysscaling}.
\end{itemize}





\section{Related Works}


\paragraph{Automating graphics editing}

As 3D graphics editing is a time-consuming and heavily specialized task, automating it can save substantial production costs and allow faster content generation. Following the remarkable success of large foundation models, the AI and graphics community has made great efforts to adapt VLM/LLMs for 3D graphics editing in tasks such as scene generation \cite{yang2024llplace3dindoorscene, hu2024scenecraft, DBLP:journals/corr/abs-2408-02211} and material editing \cite{huang2024blenderalchemy, chen2023text2tex, zsolnaifeher18gms, yang20233dstyle, richardson2023texture, DBLP:journals/tog/GuerreroHSMBM22}. VLMs, in particular, are well-suited for tasks that require visual reasoning and textual-visual relationship understanding. BlenderAlchemy~\cite{huang2024blenderalchemy}, for example, has demonstrated promising results in material editing. 3D-GPT~\cite{sun20233dgpt} shows decent language-driven 3D modeling capabilities. Both LLPlace~\cite{yang2024llplace3dindoorscene} and SceneCraft~\cite{hu2024scenecraft} leverage the spatial reasoning of VLM to generate 3D layouts. While these systems show significant success in their self-designed evaluation, the lack of a comprehensive benchmark makes it challenging to properly cross-compare their approaches even in the same task setting. 


\vspace{-1em}\paragraph{Benchmarks for graphics editing}
Following the rise of graphics editing tools, the need for a comprehensive graphics benchmark has become increasingly critical. Many existing benchmarks focus on some specific tasks of graphics editing, such as material\cite{huang2024blenderalchemy}, lighting \cite{huang2024blenderalchemy}, and the artistic quality of images \cite{liao2022, huang2024aesbench}. Other benchmarks attempt to measure semantic understanding in symbolic graphics programs \cite{yuan2024cadtalk, qiu2025can}, yet they are tailored toward their domain, like SVG programs. This makes it challenging to account for the multi-faceted nature of 3D graphics editing, which includes tasks like material editing, lighting adjustments, and complex scene composition. 

\vspace{-1em}\paragraph{Benchmarks for VLMs}
Recent efforts have focused on creating comprehensive benchmarks to evaluate VLMs across multiple dimensions. UniBench~\cite{al2025unibench}, for instance, provides an evaluation framework for nearly 60 publicly available VLMs, assessing capabilities ranging from object recognition and relational understanding to robustness. Although UniBench~\cite{al2025unibench} and similar frameworks \cite{zheng2022vlmbench, fu2024mmecomprehensiveevaluationbenchmark, hendrycks2021measuring, liu2024mmbench} provide a broader evaluation of vision-language capabilities of VLMs, they are not tailored to 3D graphics. This gap underscores the need for a specialized benchmark that addresses the unique demand of 3D graphics editing.

\section{Methods}




We aim to enable comprehensive comparison and analysis of VLM systems on 3D graphics editing tasks, which are currently bottlenecked by (1) the incomprehensive coverage of task settings with current evaluation approaches and (2) the absence of a robust 
evaluation metric due to unscalability of human evaluation and unreliability of AI-judge. To overcome these two challenges, we introduce BlenderGym—the first VLM systems benchmark tailored to 3D graphics (\cref{fig: teaser}).
BlenderGym features in (1) the comprehensive coverage of essential graphics editing tasks with support for a wide range of VLM systems, (2) quantitative evaluation with image and 3D metrics, eliminating the need for human or AI-based evaluations, and (3) support for inference time scaling experiments on graphics editing tasks.  

\begin{figure*}[h] 
\centering
\includegraphics[width=\textwidth, trim=0 82 0 90, clip]{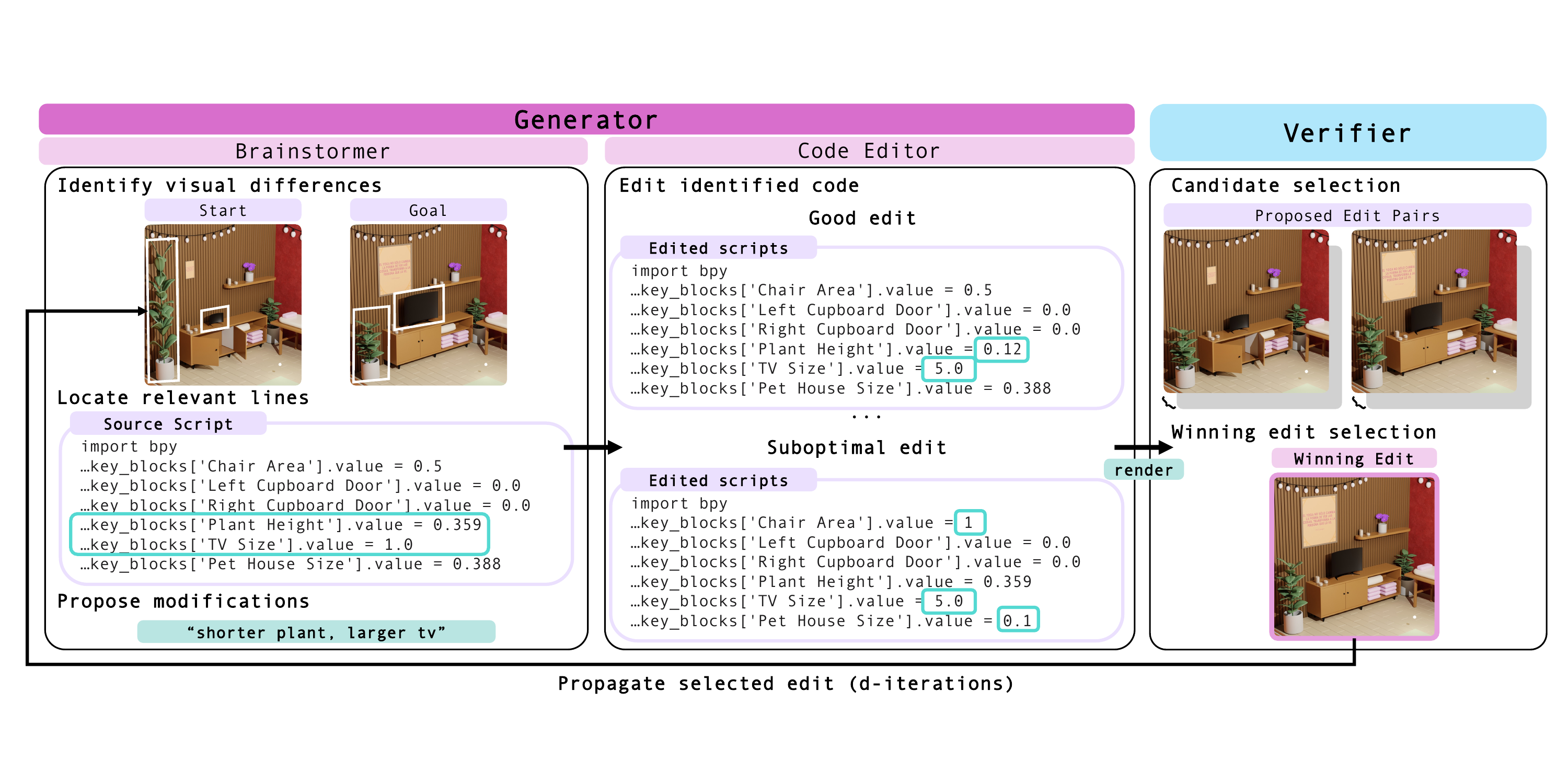}
\vspace{-20pt}
\caption{VLM system setup used by BlenderGym. It follows a generator-verifier structure, where the generator further contains a brainstormer and a code editor. The generator takes in start-goal image pairs along with Python script of the start scene, and then edits the Python script of the start scene to achieve the goal scene based on their visual differences. The verifier takes a pair of renders of proposed edits and selects a single best edit, which is propagated back to the generator for the next iteration.}
\vspace{-8pt}
\label{fig: pipeline} 
\end{figure*}

\subsection{Data Collection and Pipeline Design}
BlenderGym comprises of 245 hand-crafted Blender scenes across five key graphics editing tasks: procedural geometry editing, lighting adjustments, procedural material design, blend shape manipulation, and object placement (50/40/40/75/40 instances, respectively). Each instance in BlenderGym presents a reconstruction task from a start scene to a goal scene, where the start scene serves as the original input for the VLM system, and the goal scene represents the desired output state that the VLM system is tasked with reconstructing. Each start-goal instance includes a base Blender file of the scene setup, a pair of Python scripts that generate the start and goal scene, rendered images for both scenes, and a language description of the differences between the two scenes. Example instances of each task are demonstrated in \cref{fig: teaser}. For variety, we incorporate task instances based on Blender prototype from BlenderAlchemy~\cite{huang2024blenderalchemy}, Infinigen~\cite{raistrick2023infinite}, and Blenderkit~\cite{blenderkit}.


VLM systems run on BlenderGym in the following steps, as shown in \cref{fig: teaser}. To prepare the input data, we first run and render the start and goal Python scripts on the base Blender file. The VLM system then compares the start and goal rendered images, analyzes their visual differences, and modifies the Python script of the start scene to reconstruct the goal scene. Finally, we compute the distance between the goal scene and the VLM system output using a set of human-aligned distance metrics.


\subsubsection{Task Settings and Evaluation Metrics}

\paragraph{Procedural Material} The VLM system is expected to perceive and modify a procedural surface material, requiring an understanding of color and texture. The VLM system should edit the Python script with Blender Python API(BPY) and the Infinigen package\cite{raistrick2023infinite}. The results are evaluated using photometric loss (PL) and CLIP score \cite{radford2021learning}. 
\vspace{-2em}\paragraph{Procedural Geometry} The geometry editing task challenges VLMs in spatial and geometric reasoning capabilities, requiring them to handle variations in attributes like shape, scale, and spatial distribution. Similar to material editing, the system is expected to propose procedural edits through BPY and Infinigen. 3D geometry editing results are evaluated using Chamfer distance (CD), PL, and CLIP.
\vspace{-1em}\paragraph{Lighting Adjustments}
The lighting adjustment task involves manipulating the color, intensity, location, and orientation of the light sources to assess a VLM system’s reasoning abilities regarding the lighting environment. Notably, we include some scenes where light sources are not directly visible from the camera but can be inferred from the shadow. VLMs must edit both continuous parameters, such as RGB values, and discrete parameters, such as the type of light source (e.g. point light v.s. area light). We use PL and CLIP to evaluate the edits of lighting task.
\vspace{-1em}\paragraph{Object Placement} Object placement task requires VLM systems to perceive objects in the start scene and reposition them to their corresponding location in the goal scene. This task evaluates a VLM system's perception and reasoning of spatial information. In the input code script, we prepare various Python functions for VLM systems to modify the location of an object. They can directly set its 3D coordinates, shift its position via a vector offset, or place it relative to another object with some distance in a given direction. The distance metrics within the scene are calibrated by preserving the real-world distance scale and defining boundaries for the x, y, and z coordinates for the scene. For placement, we utilize CD, PL and CLIP to measure the quality of edits.
\vspace{-2em}\paragraph{Blend Shapes} In this task, the VLM system is responsible for shape blending -- adjusting continuous variables that control some features of an object, such as facial expression, shape, or scale. Since these blend shapes are semantically annotated with language in the code scripts, the VLM system must match the name of the blend shapes to the features they refer to and tweak their value accordingly. We leverage CD, PL and CLIP to measure the resulting edits as blend shapes usually involve 3D structures. 
\begin{figure*}[t]
\begin{center}
\includegraphics[width=\linewidth, trim=0 90 0 82, clip]{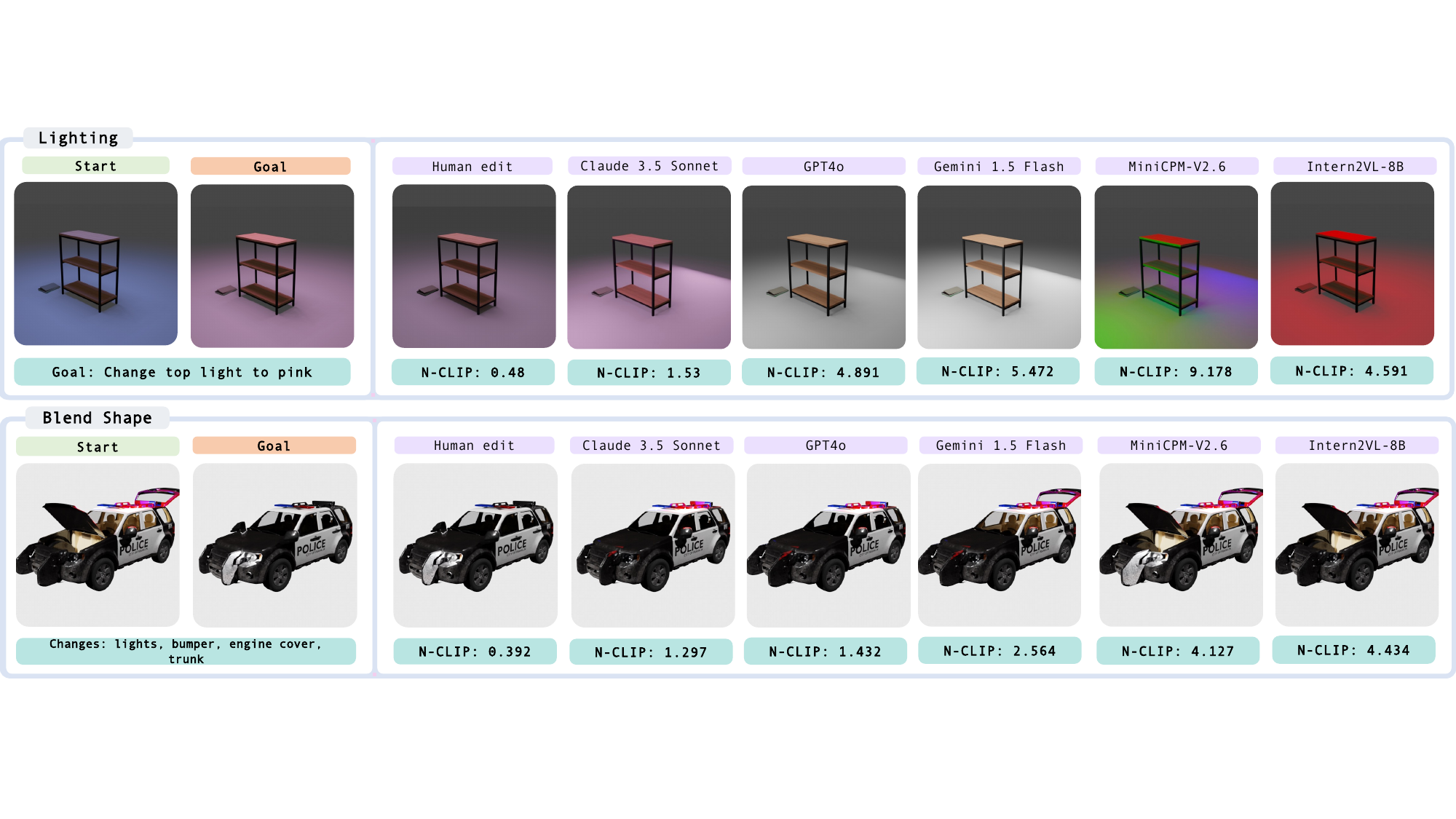}
\end{center}

\vspace{-1em}

    \caption{Examples of VLM/human generated outputs on lighting and blend shape tasks. Even powerful closed-source VLMs fail to generate lighting settings with accurate colors. In the blend shape task above, only Claude 3.5 Sonnet correctly captures the overall appearance of the police car but still misses the difference of the front light. We report N-CLIP ($10^{-3}$) for all edits above as a calibration.}
    \label{fig: quali_scenes}

\end{figure*}

Note that despite our choice of name for blend shapes and our usage of Infinigen\cite{raistrick2023infinite} as wrapper functions for Blender-Python API in procedural material and geometry editing tasks, BlenderGym users can experiment with other representations of the start and goal Blender scenes.   

For evaluation, we use negative CLIP similarity (N-CLIP), defined as $(1-\text{CLIP Score})$ to clip the CLIP scores to positive values and reverse its trend, keeping it consistent with photometric loss and Chamfer distance such that a lower metric value indicates greater similarity to the goal.

\subsubsection{Camera Viewpoint Selection}
\label{subsubsec: camera viewpoint}
Occluded or incomplete render views of the 3D scene can introduce bias for VLM edits. Therefore, for all the 3D-related tasks, we manually select and fix at least three render views for each start-goal scene pair, ensuring that (1) at least one comprehensive view, in which most objects are exposed, is included and that (2) all objects are captured by at least one camera view. A more detailed definition and examples of comprehensive views can be found in \cref{sec: camera view}. For VLM input, we use two camera views, with at least one being a comprehensive view. For evaluation, we use all the views. Note that since we have ensured that evaluation views cover every object in the scene, BlenderGym users can experiment with different choices of camera position or visual representation for VLM system input.




\subsection{Human Baseline}
\label{subsec: human_in_the_loop}
We demonstrate the solvability and validity of BlenderGym and show the performance gap between VLMs and humans by engaging 10 average Blender users to complete approximately 20\% of all task instances in BlenderGym. We restrict the edit time of human users to at most 8 minutes per instance, which matches the average time taken by a VLM system. We then evaluate human edits using the same metrics as for VLMs. The results of human Blender users are shown in \cref{tab: selected table} along with that of VLM systems.

\section{Experiment}
In this section, we provide a comprehensive analysis of VLM system performance on BlenderGym, present human baselines, and explore methods to enhance the editing capabilities of VLM systems.

\begin{table*}[ht]
\centering
\resizebox{\textwidth}{!}{%
\begin{tabular}{@{}lccccccccccccc@{}}
\toprule
\multicolumn{1}{c}{} & \multicolumn{3}{c}{Blend Shape} & \multicolumn{3}{c}{Placement} & \multicolumn{3}{c}{Geometry} & \multicolumn{2}{c}{Lighting} & \multicolumn{2}{c}{Material} \\ 
\cmidrule(lr){2-4} \cmidrule(lr){5-7} \cmidrule(lr){8-10} \cmidrule(lr){11-12} \cmidrule(lr){13-14}
Model & PL $\downarrow$ & N-CLIP $\downarrow$ & CD $\downarrow$ & PL $\downarrow$ & N-CLIP $\downarrow$ & CD $\downarrow$ & PL $\downarrow$ & N-CLIP $\downarrow$ & CD $\downarrow$ & PL $\downarrow$ & N-CLIP $\downarrow$ & PL $\downarrow$ & N-CLIP $\downarrow$ \\ \midrule
GPT-4o & \textbf{9.140} & \textbf{20.47} & \textbf{0.904} & 11.89 & \textbf{30.38} & 11.22 & \textbf{6.747} & 8.561 & 1.192 & \textbf{2.410} & 2.398&  \textbf{3.653}& 8.942 \\ 
Claude-3.5-Sonnet & 12.79& 27.96 & 1.962 & 13.19 & 51.76 & 11.29 & 10.81 &  13.04 &  1.452 & 2.897 & 4.049 &  5.769 & 11.44 \\ 
GPT-4-Turbo & 15.21 & 26.15 & 1.927& 12.21 & 37.57& 12.80 &  8.160 & 10.92 & \textbf{1.120} & 2.723 & 3.912&  5.424 &  \textbf{8.812}\\ 
Claude-3-Haiku& 13.62 & 29.72 & 2.563 & 14.78 & 44.10 & 12.13 & 10.15 &  12.51 &  1.362 & 3.712 & 4.824 &  5.960 &  11.61 \\ 
Gemini-1.5-flash & 23.18 & 30.47 & 2.412 &  \textbf{10.94} &  45.34 & \textbf{8.324} &  9.443 & 10.49 & 1.323 & 3.514 & 5.688 &  6.364 & 10.42 \\ 
Qwen2-vl-7b & 16.78 & 29.22 & 2.123 & 15.31 & 41.12&  14.21 &  --&  --&  --& 2.985 & \textbf{2.225} &  --&  --\\ 
Qwen-Llama & 14.32& 28.23 & 2.012 & 14.65 &  34.93 &  12.41 &  13.97 & 14.13 &  1.673 & 3.173 & 3.998 &  --&  --\\ 
Phi-3.5-vision & 12.51 & 24.14 & 2.012 &  -- &  -- &  -- & -- &  -- &  -- &  3.127 & 6.012 &  --&  --\\ 
Phi-Llama & 12.13 & 24.77& 1.826 & 14.61 & 35.61 & 12.61 & 9.818&  11.92 & 1.471 & 3.621 & 6.895 &  --&  --\\ 
MiniCPM-V-2.6 & 13.86 & 29.92 & 1.997 & 11.99 & 31.69 & 12.62&  7.127 & \textbf{8.542} & 1.229 & 3.829 & 6.124 &  -- &  -- \\ 
MiniCPM-Llama & 13.76& 27.21 & 1.882 & 12.74 & 31.72 & 15.81 &  9.561 & 11.47 & 1.569 & 3.725 & 6.090 & 7.152 &  12.14\\ 
InternVL2-8b & 12.69 & 29.09 & 1.920 & 14.71 & 35.92 & 17.22 &  --&  --&  --& 3.920 & 6.825 &  --&  --\\ 
Intern-Llama & 11.80 & 23.83 & 1.861 & 16.15 & 37.23 & 18.22 &  13.70 & 14.44 & 1.578 & 3.825 & 6.152 &  --&  --\\ 
\midrule
\textbf{Human}& \textbf{0.934}& \textbf{9.12}& \textbf{0.399}& \textbf{0.423}& \textbf{13.34} & \textbf{1.532} & \textbf{1.269} & \textbf{2.434} & \textbf{0.334} & \textbf{1.239} & \textbf{1.632} & \textbf{0.629} & \textbf{3.043}  \\

\bottomrule
\end{tabular}
}
\caption{Quantitative evaluation of VLM system edits. The magnitude of photometric loss(PL) and negative-clip(N-CLIP) is scaled to $10^{-3}$ for blend shape and placement tasks, and to $10^{-2}$ for the remaining tasks. We report unscaled Chamfer Distance(CD). Values marked with ``-'' indicate that the VLM system was unable to generate any executable code script in more than $75\%$ of the instances for that task due to the complexity in the code scripts of these two tasks.  
}
\vspace{-10pt}
\label{tab: selected table}
\end{table*}

\subsection{VLM System Setup}

\label{subsec: system design}
We base our VLM setup on the BlenderAlchemy~\cite{huang2024blenderalchemy} multi-agent system, as it is the current SOTA pipeline for multiple graphics editing tasks and supports seamless integration of various VLMs. Our implementation leverages a generator-verifier structure shown in \cref{fig: pipeline}, with two hyper-parameters: depth $d$ and breadth $b$. 

\vspace{-1em}\paragraph{Generator} The generator has two key model components: the brainstormer and the code editor. The brainstormer identifies the most significant visual difference between the rendered images of the start and goal scenes, examines the Python script of the start scene to locate relevant lines, and proposes instructions to edit the start script. These instructions may take the form of natural language or Python code. The code editor then incorporates the instructions into an edited Python script. This brainstormer-code-editor structure is repeated independently for a breadth $b$ times to derive a list of $b$ parallel Python script proposals. 
\vspace{-1em}\paragraph{Verifier} The verifier receives $b$ proposals from the generator and splits them into $\frac{b}{2}$ pairs in sequential order. It then compares the render images of the candidates with the goal render and selects the one closest to the goal. The selected edits from all pairs are then aggregated into a list of $\frac{b}{2}$ winner candidates, on which we perform pair-wise selection iteratively until only one candidate remains as the winner.

One generator-verifier pair forms an edit iteration that incrementally makes the input scene more closely resemble the goal scene. For the sake of consistency, our setup uses the \textbf{same VLM} for both the generator and the verifier. The output of each iteration is propagated as the input scene for the subsequent iteration. We repeat such iteration for a depth $d$ times, with the final output from the last iteration representing the result of the entire pipeline. The overall structure resembles a $d \times b$ tree, where each iteration prunes suboptimal edits and progressively refines the best candidate. For this benchmark, we set $d=3$ and $b=4$. Prompts used for each agent are provided in \cref{sec: prompts}. Note that despite our usage, BlenderGym is not limited to BlenderAlchemy, but supports easy plug-and-play of any VLM system. Other input information, such as prompt and camera views, can also be customized by the user.

\subsection{Benchmark Results}
\label{subsec: benchmark_results}
We benchmark 13 closed- and open-source VLMs, including GPT-4o~\cite{2023GPT4}, Claude 3.5 Sonnet~\cite{claude3_model_card}, Gemini 1.5 Flash~\cite{team2023gemini}, InternVL2-8B~\cite{chen2024internvl}, and Qwen2-VL-7B-Instruct-AWQ~\cite{bai2023qwen}. For VLMs that do not accept text-only input, we prompt them with a random noise image to meet the format requirements. As discussed in \cref{subsec: human_in_the_loop}, we also evaluate the performance of 10 human Blender users as a baseline to compare VLM performance with. We calibrate the N-CLIP metric in \cref{fig: quali_scenes} and other metrics in \cref{sec: calibration} to illustrate these metric values visually.

\cref{tab: selected table} shows quantitative results on VLM and human performance in all tasks. Throughout the experiments, we observe a substantial performance gap between human Blender users and VLM systems (\cref{tab: selected table} \& \cref{fig: quali_scenes}), indicating that graphics editing remains an unsolved challenge for VLMs. Furthermore, performance varies significantly between different VLM systems, indicating that different VLM systems might be better suited for different tasks.

\subsubsection{Generation Analysis}
\label{subsubsec:generation_analysis}
VLM generators remain imperfect, as illustrated by the qualitative examples in \cref{fig: quali_scenes}. Among the failure cases, we observe a few factors that significantly contribute to suboptimal edits. We attach examples of each in \cref{sec: genfailure}.
\vspace{-1em}\paragraph{Generator fails to capture subtle visual differences.} The VLM sometimes struggles to detect visual differences between start and goal scenes, essential for fine-tuning graphical edits. 
\vspace{-1em}\paragraph{Generator fails to produce an executable Blender Python script.}  VLM systems perform particularly poorly in procedural material and geometry editing, as these tasks require wiring of the nodes of the shader editor and geometry editor in Blender Python API through code editing. All open-source VLM systems, with the exception of MiniCPM, fail to produce any executable code scripts for more than 75\% of task instances in geometry editing task. To address this, we incorporate Llama3.1-8b\cite{llama31}, known for its strong performance on code editing benchmark, as the code editor model for the generator described in \cref{subsec: system design}. Performance of Llama-enhanced VLMs is marked with a suffix Llama in \cref{tab: selected table}. While Llama indeed enables Phi, Qwen, and Intern to complete more than 75\% of geometry tasks, this combination does not necessarily lead to stronger performance for all task settings.
\vspace{-1em}\paragraph{Generator provides irrelevant code changes to reflect the visual differences spotted.} Even if the resulting proposal is executable in Blender, it may not reflect the visual differences indicated by the brainstormer. This problem is particularly common with complex Blender Python scripts such as those of procedural material or geometry editing tasks, which usually consist of more than 80 lines of code.


\subsubsection{Verification Analysis}
\label{subsubsec:verification_analysis}

The verifier sometimes fails to distinguish between desirable and suboptimal edits. Certain VLMs, such as Qwen~\cite{bai2023qwen}, consistently favor the second edit candidate in the pair, regardless of how we permute them. We provide example failure cases of VLM verifier in \cref{sec: verfailure}.

\begin{figure}
\centering
\includegraphics[width=.8\linewidth]{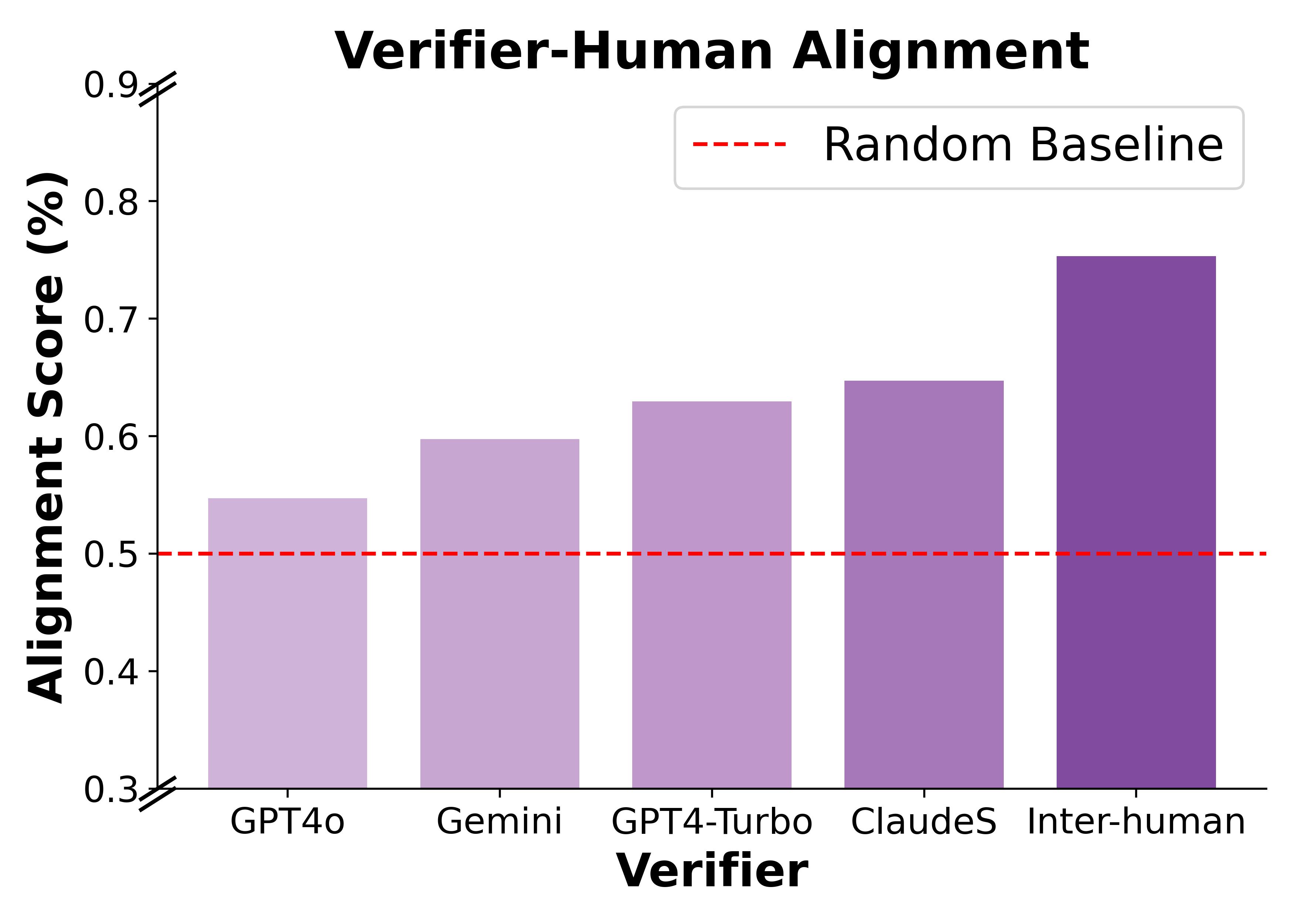}
\vspace{-10pt}
\caption{Human-VLM and inter-human verifier alignment rate. All models perform above the random baseline (0.5) yet differ notably, with even the highest-aligned model, Claude-3.5-Sonnet (0.66), falling short of inter-human alignment (0.79).}
\label{fig: alignment}
\vspace{-16pt}
\end{figure}

We measure human-VLM verifier alignment to ground our verifier analysis more quantitatively. For each of the five tasks, we select one instance and collect all the pair-wise selection decisions that the VLM verifier made throughout the inference. We collect 7,950 pair-wise judgments from 50 participants. For each pair-wise selection, VLM and human verifiers are asked to choose the edit render closest to the goal render. We compute the alignment rate between two verifiers by calculating the ratio of the number of aligned pair-wise selections over the total number of pairs. We compute the alignment rate between human and GPT-4o, GPT-4 Turbo, Claude 3.5 Sonnet, and Gemini 1.5 Flash, and also measure inter-human alignment. Results in \cref{fig: alignment} reveal that human verifiers achieve a significantly higher alignment rate compared to VLMs, highlighting considerable room for improvement in VLM verifier performance.

\setlength{\intextsep}{0pt}
\setlength{\columnsep}{10pt}

\subsection{Verifier Scaling}
\label{subsec:verifier_scaling}
To complement recent findings on inference scaling of VLM/LLM generation\cite{brown2024largelanguagemonkeysscaling, snell2025scaling}, we explore improving the verifier that guides the generation by selecting desirable edits and pruning suboptimal ones. \citet{brown2024largelanguagemonkeysscaling} demonstrates that, assuming the presence of an oracle verifier that always identifies the best answer, scaling open-source models for generation can outperform closed-source models in coding and math problems. However, when this impractical oracle verifier is replaced by practical verifiers like majority voting or reward models, the accuracy of open-source LLM outputs plateaus quickly below that of closed-source models.
Another verifier choice explored is the task-specific process-based reward model (PRM), which has been shown to be effective for math problems by ~\citet{lightman2024lets} and ~\citet{snell2025scaling}, but it requires additional training on task-specific data. Therefore, finding cost-effective verifiers that do not require domain-specific data remains an open challenge. To fill this gap, we propose \textbf{inference time scaling for VLM verifier}, the scaling of VLM as an answer selector, as an alternative path. This quantitative experiment is enabled by BlenderGym and the scalable visual state evaluator structure within our pipeline in \cref{fig: pipeline}.

We find that (1) similar to generation, VLM verifiers used for guiding generation also benefit from inference scaling, and that (2) the performance of scaled open-source VLM verifiers can exceed that of closed-source ones.

\subsubsection{Experiment Details}
\label{subsubsec: verifier_scaling_details}
To scale up the verifier, we introduce a reselection parameter $k$ into the verifier structure (\cref{alg: scaled}). We run the vanilla verification process $k$ times to generate a list of $k$ winners, each time shuffling the candidate list. Then, we select a single final winner from the winner list.

\begin{algorithm}[t]
\caption{Scaled Verification}
\textbf{Function} ScaledSelect(List of proposal candidates $\{S_1, S_2, \dots, S_n\}$, Verifier $V$, Reselection parameter $k$): \\
\textbf{Initialize} \textit{winner\_list} as an empty list. \\
\textbf{for} $i = 1$ to $k$ \textbf{do} \\
\quad $\{S_1', S_2', .., S_n'\} \gets \text{shuffle} (\{S_1, S_2, .., S_n\})$ \\
\quad $winner \gets \text{SingleSelect}(\{S_1', S_2', .., S_n'\}, V)$ \\
\quad Append $winner$ to \textit{winner\_list}. \\
\textbf{end for} \\
$final\_winner \gets \text{SingleSelect}(\text{winner\_list}, V)$ \\
\textbf{return} $final\_winner$
\label{alg: scaled}
\end{algorithm}

\begin{figure}[t] 
    \centering
    \vspace{-5pt}
    \includegraphics[width=\linewidth, trim=250 0 470 0, clip]{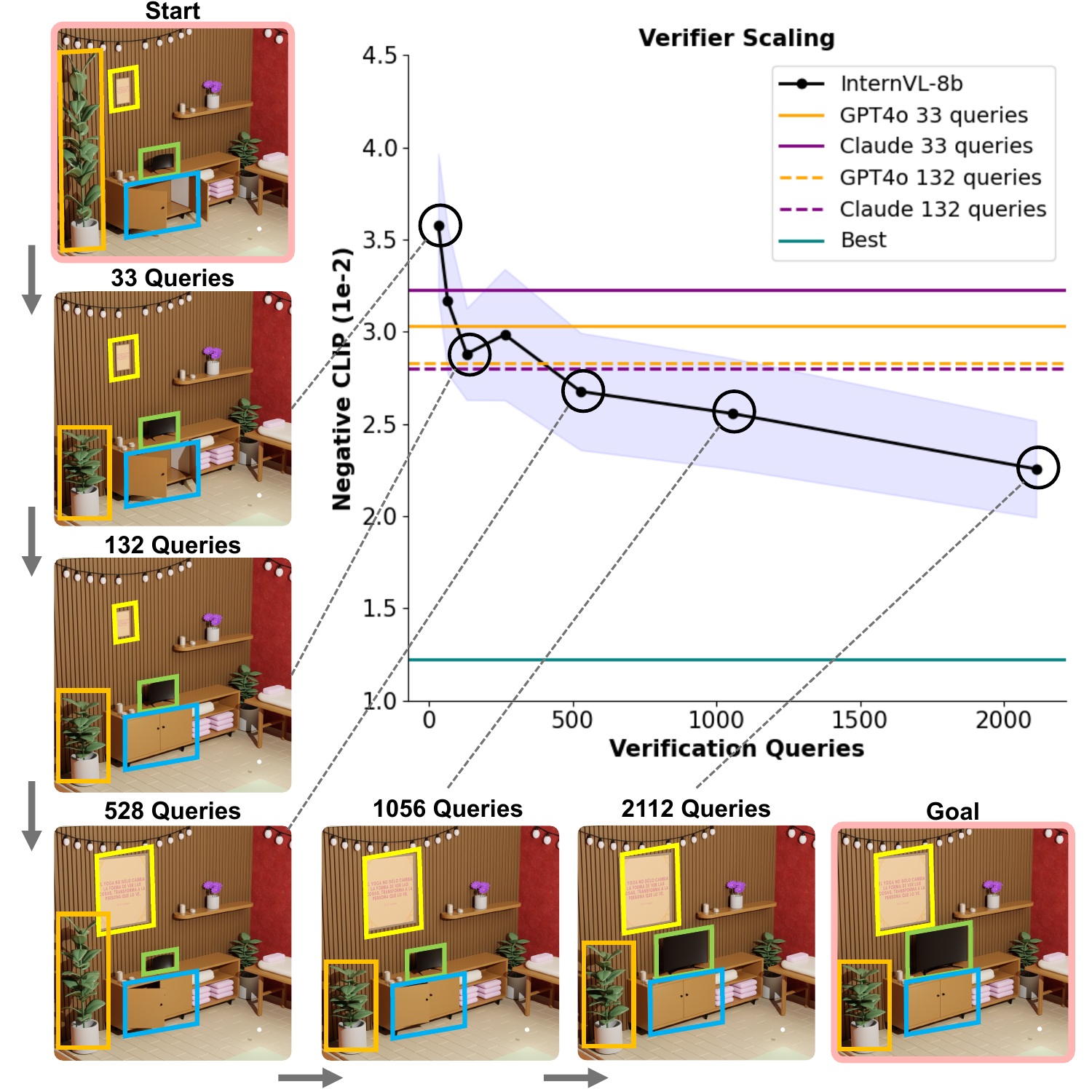}
    \vspace{-15pt}
    \caption{Verifier scaling results with InternVL2-8B, Claude3.5 Sonnet, and GPT4o. We show that increasing verifier queries brings the selected edit closer to the goal. Bounding boxes of all object instances of interest are shown.}
    \label{fig: verifier_scaling} 
    \vspace{-10pt}
\end{figure}

With the scaling mechanism defined, we select 16 instances of blend shape task and use Claude 3.5 Sonnet to generate 32 candidate edits (4x8 tree) for each blend shape instance. We fix these 32-candidate trees and run ScaledSelect (\cref{alg: scaled}) with InternVL2-8B to select the best edit from the trees. We vary the reselection parameter $k$ in 1, 2, 4, 8, 16, 32, and 64. \cref{fig: verifier_scaling} presents the metric scores and qualitative examples of the finally selected edit for different levels of verification queries. We also report smaller-scale scaling results for GPT-4o and Claude 3.5 Sonnet.  

Three key findings emerge from this experiment. (1) As inference compute increases for verification, the values of all three metrics of the selected edits decrease, indicating improved verifier selection. 
(2) Scaled InternVL2-8B outperform unscaled or slightly scaled GPT-4o and Claude 3.5 Sonnet in all three metrics. (3) Closed-source VLM systems also benefit from scaling. We conclude that for graphics editing, inference scaling is effective not only for generation but also for VLM verification that guides the generation. These findings encourage further exploration of the scalability of verification.

\subsection{Allocation of Compute}
\label{subsec:allocation}
With the knowledge that VLM generation and verification both benefit from inference scaling, we further explore whether an optimal distribution of inference compute between generation and verification exists. We experiment with different ratios of verification queries over total queries (VeriRatio) and find that (1) the choice of VeriRatio significantly impacts the VLM performance and that (2) the optimal VeriRatio varies with the level of total compute — more total compute benefits from a higher share of verification.

\subsubsection{Experiment Details}
To measure the impact of compute allocation on VLM system performance, we fix three VeriRatios and run our setup across different total compute levels, measured by the total amount of queries. We use InternLlama with total queries varying from 12 to 245 and set reselection parameter $k$ (defined in \cref{subsubsec: verifier_scaling_details}) to 1, 3, and 5, corresponding to VeriRatios of 0.33, 0.62, and 0.73, respectively.

\begin{figure}[t] 
    \centering
    \includegraphics[width=\linewidth, trim=0 0 0 0, clip]{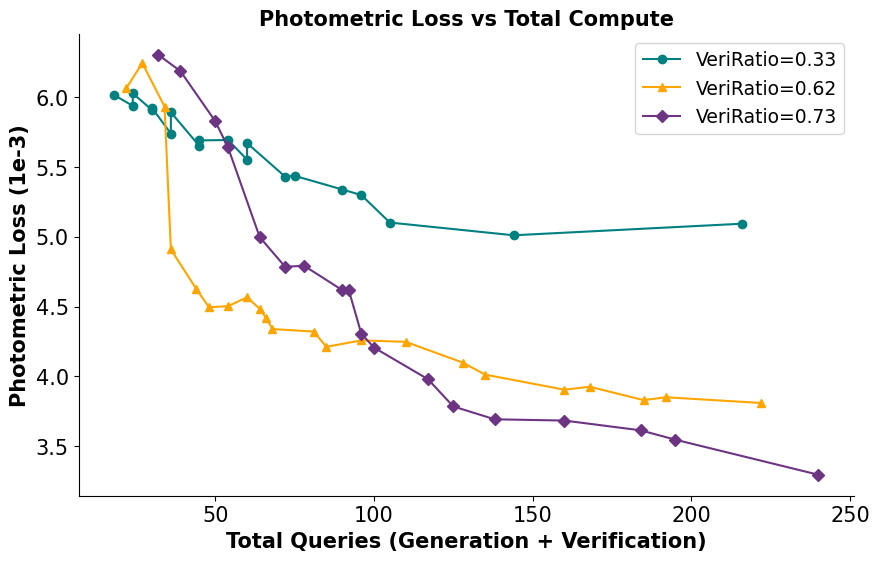}
    \vspace{-20pt}
    \caption{The impact of compute allocation on VLM system performance for 16 blend shape task instances. We show photometric loss at VeriRatios of 0.73, 0.62, and 0.33. We observe that with fewer compute, the generation process dominates the performance of the whole pipeline, while with a large compute budget, increasing verifier compute is more effective. The compute unit is query as every 100 generation/verification queries incur a similar cost of 0.45 USD.}
    \label{fig: allocation} 
    \vspace{-10pt}
\end{figure}

Note that rather than selecting from a fixed candidate pool repeatedly as in \cref{subsec:verifier_scaling}, the reselection algorithm here is integrated into the pipeline (\cref{fig: pipeline}) to allow a scaled real-time pruning of suboptimal edits at each iteration. We experiment on 16 blend shape tasks (see \cref{fig: allocation}) and 15 lighting tasks (see \cref{sec: veriscaling}). 

\cref{fig: allocation} demonstrates that the allocation of inference compute significantly impacts VLM system performance. We observe from \cref{tab:veriratio} that as the total compute increases, a higher VeriRatio leads to a better performance. This suggests that when total inference compute is limited, generating a variety of candidates benefits more than stronger verification. However, when the inference budget is sufficiently high, refining a relatively smaller batch of candidate edits is more helpful than generating a larger set of potentially suboptimal candidates. This points to a broader insight about the importance of ``deliberative'' compute, or strategically allocated compute, in graphics editing: improvement in performance demands a shift from sheer generation capacity to a propose-verify-improve iterative workflow.


\begin{table}[h!]
\vspace{10pt}
\centering
\begin{tabular}{|c|c|c|c|c|}
\hline
 & \multicolumn{4}{c|}{Total Number of Queries} \\ \hline
 & $<30$ & 30-60 & 60-100 & $>100$  \\ \hline
VeriRatio=0.33          & \textbf{1}          & 2          & 3 & 3       \\ \hline
VeriRatio=0.62          & 2          & \textbf{1}          & \textbf{1}    & 2       \\ \hline
VeriRatio=0.73         & 3          & 3          & 2     & \textbf{1}      \\ \hline

\end{tabular}
\caption{Performance rank of InternLlama with different VeriRatio across total compute levels. We observe that as the total compute, measured by total number of queries, increases, larger VeriRatio leads to better performance.}
\label{tab:veriratio}

\end{table}






\section{Conclusion and Discussion}
In this work, we introduce BlenderGym, a comprehensive 3D graphics benchmark that tasks VLM systems with 3D scene reconstruction through code editing. Our experiments reveal that graphics editing remains unsolved for VLM systems. Furthermore, we demonstrate on BlenderGym that scaling verifiers can improve the VLM system's performance and that the more inference compute allowed, the higher proportion of it should be allocated to verification to optimize the system performance. 

\vspace{-1em}\paragraph{Limitations and future works} 
While BlenderGym provides a robust foundation, further improvements are possible in the following areas. (1) The coverage and scale of BlenderGym can be expanded to more graphics editing tasks, such as object sculpting, camera-view adjustment, and animation creation with more data instances. (2) VLM-human verifier alignment and human Blender generator experiment can be enhanced by increasing the number of human annotators beyond the current scale. (3) More approaches can be explored to improve verifier performance, as they sometimes still fail to select the best edit despite being scaled. Exploring more advanced verifier pipelines or scaling up closed-source VLMs may offer further benefits.

\vspace{-1em}\paragraph{Societal Impacts} The automation of 3D graphics editing introduces complex socio-technological challenges, as graphics editing is a creative human-driven effort. The integration of VLMs into the system is costly and can reduce opportunities for human involvement due to the replacement of creative toolsets. Nonetheless, BlenderGym allows us to systematically study VLM systems and bridge the gap between graphics editing and the development of foundational models. Future work in automating the editing workflow must carefully consider these risks and prioritize mitigating them.

\vspace{-1em}\paragraph{Acknowledgements} We acknowledge the support of ARL grant W911NF-21-2-0104 and a Vannevar Bush Faculty Fellowship. Additionally, we would like to thank Yangjun Ruan, Haiwen (Haven) Feng, Jordan Juravsky, and all our reviewers for their feedback on the paper revisions.


\onecolumn
\begin{center}
    \vspace{-18pt}
    {\large \textbf{BlenderGym: Benchmarking Foundational Model Systems for Graphics Editing}} \\[1em]
    {\normalsize Supplementary Material}
\end{center}
\vspace{3pt}

\appendix

\renewcommand{\thesection}{S\arabic{section}}

\setcounter{table}{0}
\renewcommand{\thetable}{A\arabic{table}}

\section*{Appendix Overview}
In ~\cref{sec: veriscaling}, we extend the verifier scaling experiments to broader tasks. In ~\cref{sec: genfailure}, we analyze generator failure examples. In ~\cref{sec: verfailure}, we show a verifier's decision process for a task instance and offer a cross-model verification comparison on the same set of instances. In ~\cref{sec: calibration}, we provide a calibrated interpretation of evaluation metrics and analyze their limitations. In \cref{sec: camera view}, we show the reasoning behind the camera-view selection. Finally, in \cref{sec: prompts}, we provide all the prompts used by the generator and verifier.

\section{Verifier Scaling}
\label{sec: veriscaling}
\begin{figure}[H]
\centering
\includegraphics[width=\linewidth]{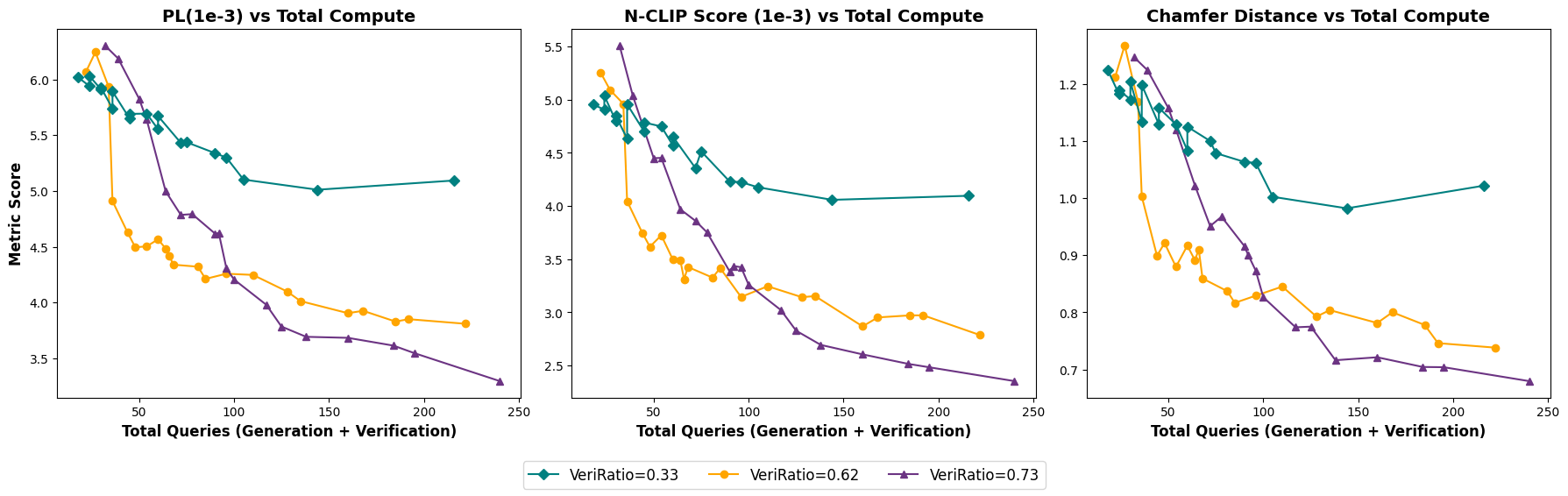}
\vspace{-20pt}
\caption{Impact of compute allocation in all three metrics on \textbf{blend shape manipulation}  task.}
\label{fig: all_metrics_shapekey}
\end{figure}

\begin{figure}[H]
\centering
\vspace{18pt}
\includegraphics[width=0.68\linewidth]{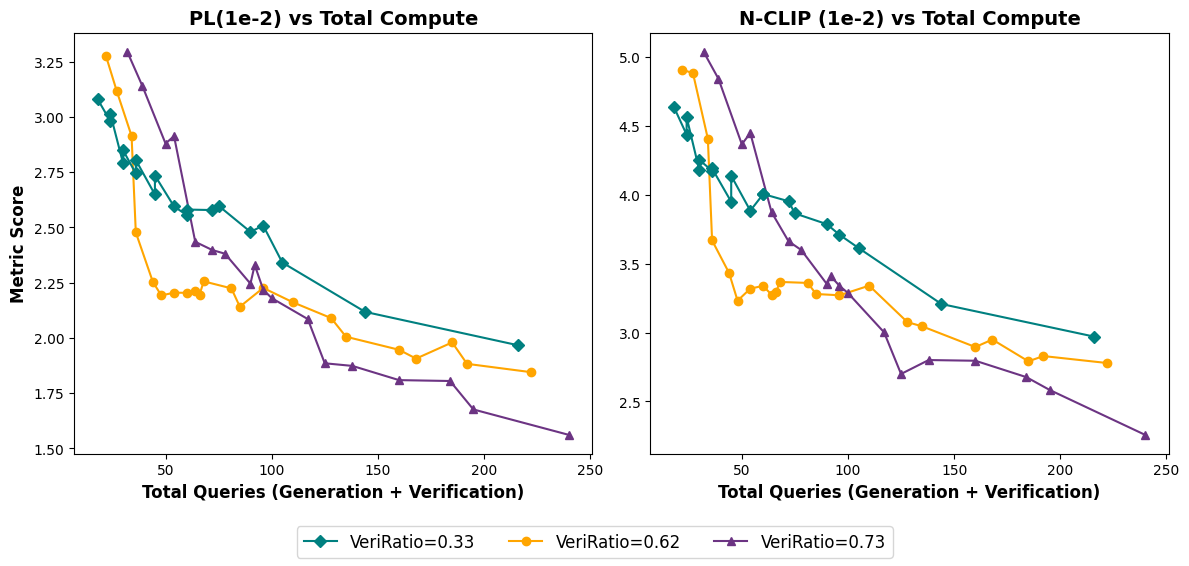}
\vspace{-8pt}
\caption{Impact of compute allocation in both PL and N-CLIP metrics on \textbf{lighting adjustment} task.}
\label{fig: all_metrics_lighting}
\end{figure}

To consolidate our findings on strategic compute allocation between verification and generation, we (1) plot the N-CLIP score and Chamfer distance for blend shape (\cref{fig: all_metrics_shapekey}) and (2) extend our experiments to 15 lighting task instances (\cref{fig: all_metrics_lighting}).

Our results demonstrate that with increased compute, VLM systems with higher verification ratio \textbf{consistently} outperform those with lower verification ratio. However, the size of this performance gap varies across tasks. As shown in \cref{fig: all_metrics_shapekey} and \cref{fig: all_metrics_lighting}, the performance gap between higher and lower verification ratios is smaller for lighting than for blend shape tasks. Our interpretation is that this gap is \textbf{positively related} to the difficulty of verification for the task --Lighting involves assessing more prominent factors like light intensity and color and, therefore, is easier for verification. In contrast, the blend shape manipulation task requires detecting more subtle and continuous changes, posing a significantly greater challenge.

We directly use the summation of the generation and verification queries as total queries since they incur a similar cost. 

\vspace{8pt}
\setlength{\intextsep}{0pt}
\setlength{\columnsep}{10pt}

\setlength{\intextsep}{0pt}
\setlength{\columnsep}{10pt}

\clearpage

\section{Generator Failure Cases}
\label{sec: genfailure}
\begin{figure}
\centering
\includegraphics[width=.95\linewidth, page=1]{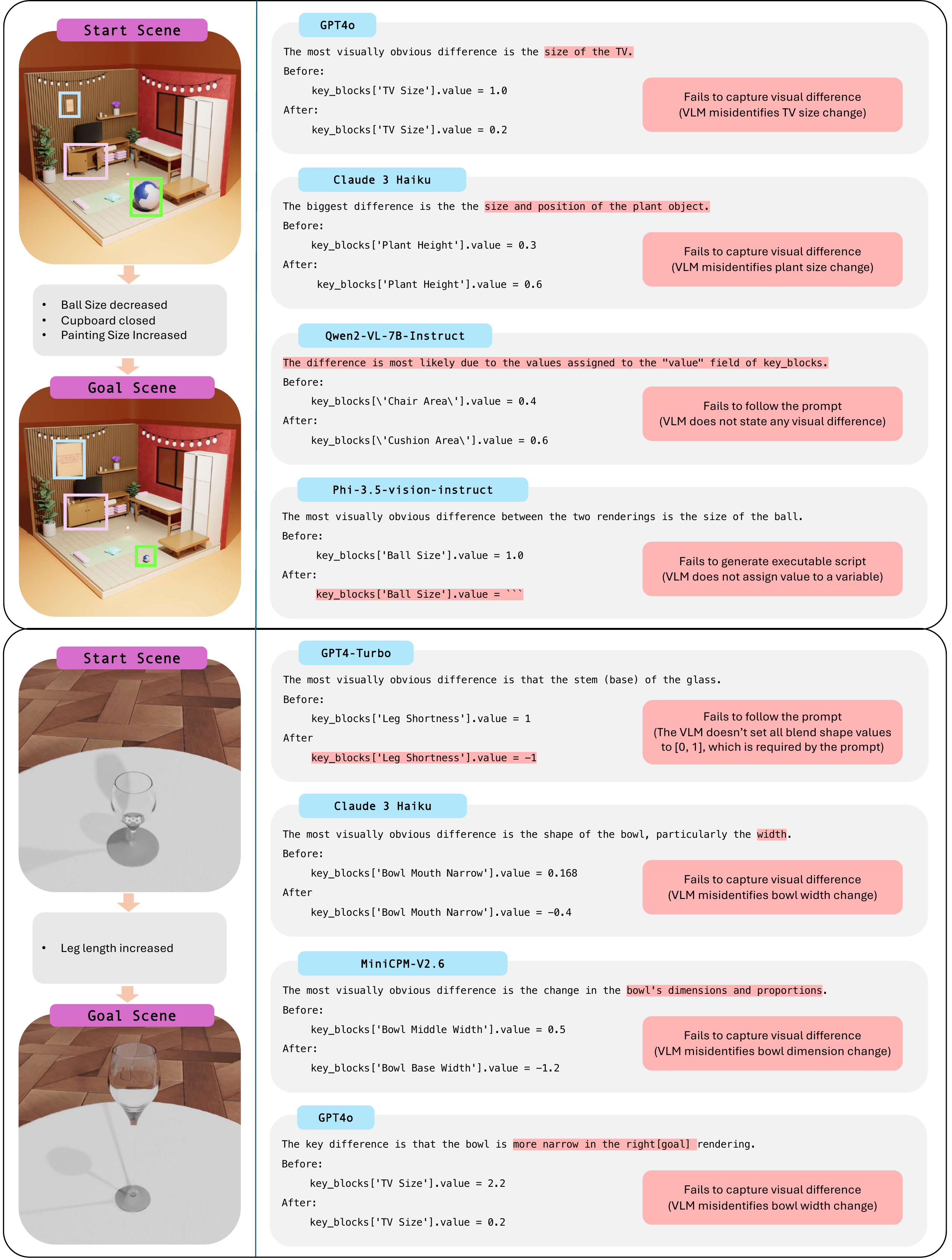}
\caption{Examples of generator failure for blend shape manipulation. We present the most visually obvious difference observed by the VLM, the code change proposed, and a failure analysis.}
\label{fig: gen_fail_1}
\end{figure}

\begin{figure}
\centering
\includegraphics[width=.95\linewidth, page=2]{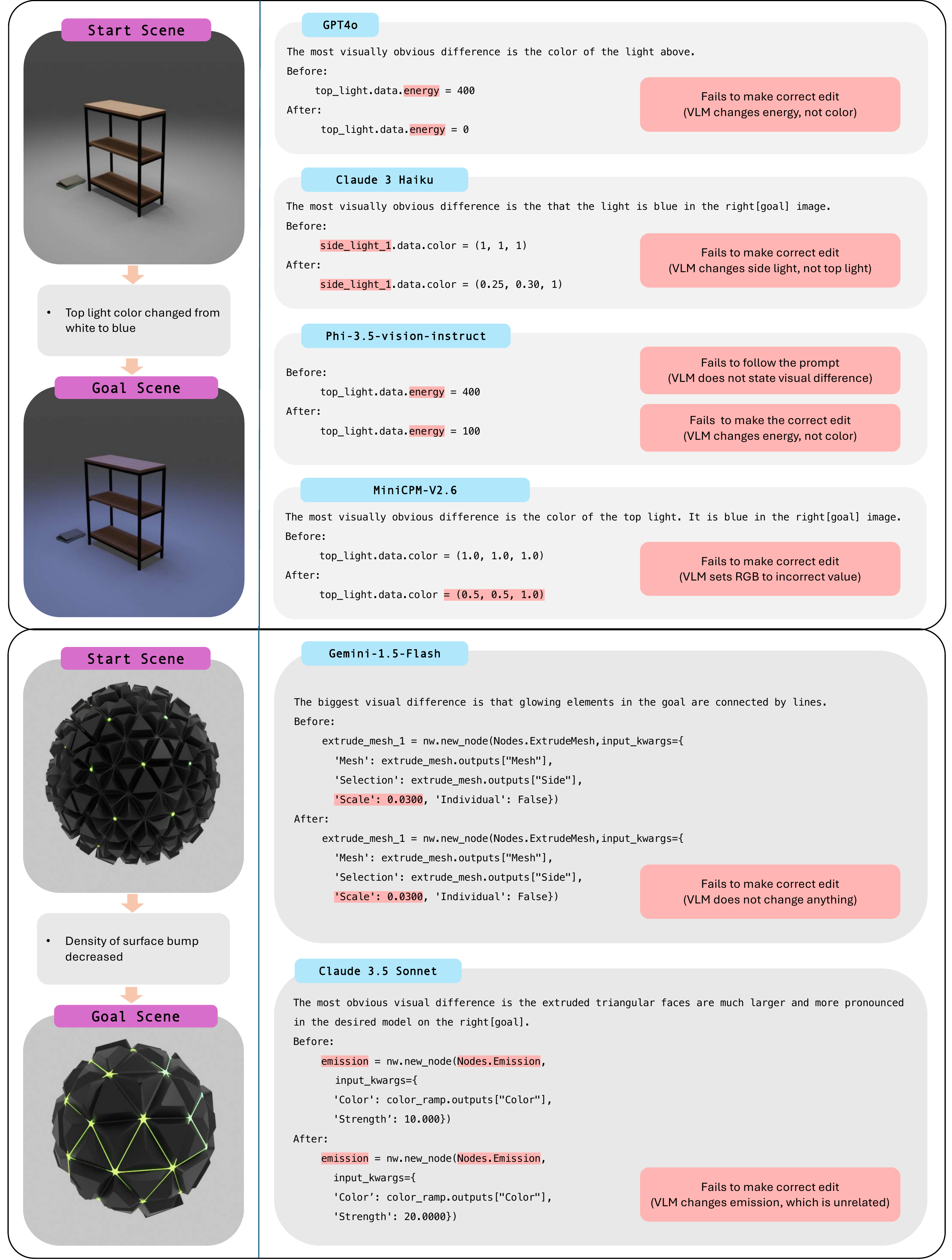}
\caption{Examples of generator failure for lighting and procedural geometry. We present the most visually obvious difference observed by the VLM, the code change proposed, and a failure analysis.}
\label{fig: gen_fail_2}
\end{figure}

\begin{figure}
\centering
\includegraphics[width=.95\linewidth, page=3]{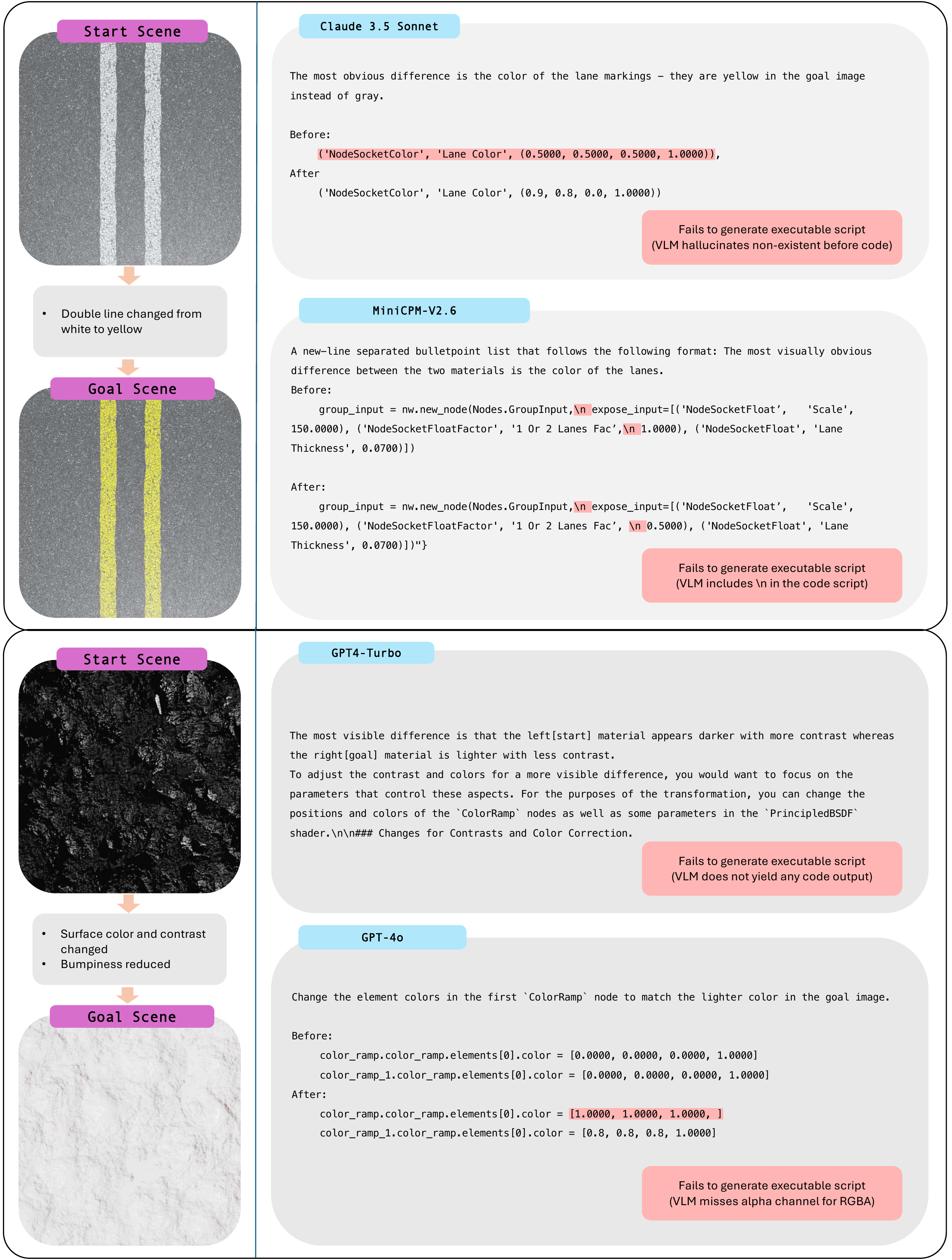}
\caption{Examples of generator failure for procedural material editing. We present the most visually obvious difference observed by the VLM, the code change proposed, and a failure analysis.}
\label{fig: gen_fail_3}
\end{figure}

Despite their capabilities, VLM generators exhibit the following common failures, as shown in \cref{fig: gen_fail_1}, \cref{fig: gen_fail_2}, and \cref{fig: gen_fail_3}:

\vspace{-13pt}\paragraph{Failure to capture subtle visual differences.} This issue arises mainly because the VLM generator often hallucinates non-existent visual differences between images rather than identifying actual discrepancies. This is a well-known limitation of VLMs and remains challenging to address. To mitigate this, we employ chain-of-thought (CoT) prompting, instructing the VLM to begin by analyzing visual differences and ignoring the code script. Details on our CoT implementation can be found in \cref{sec: prompts}. However, the generator still occasionally disregards the CoT prompt, prematurely suggesting code changes instead of reasoning step-by-step, disrupting the intended stable reasoning process.

\vspace{-13pt}\paragraph{Failure to produce executable Blender Python scripts.} Tasks like procedural material and geometry editing present significant challenges in generating executable code that reflects intended changes. These failures often stem from syntax errors, incompatibility with the Blender-Python API, or the inability to effectively incorporate the visual differences identified by the VLM, ultimately resulting in incorrect modifications.

\section{Verifier Failure Cases}
\label{sec: verfailure}
\begin{figure}[H]
\centering
\includegraphics[width=\linewidth]{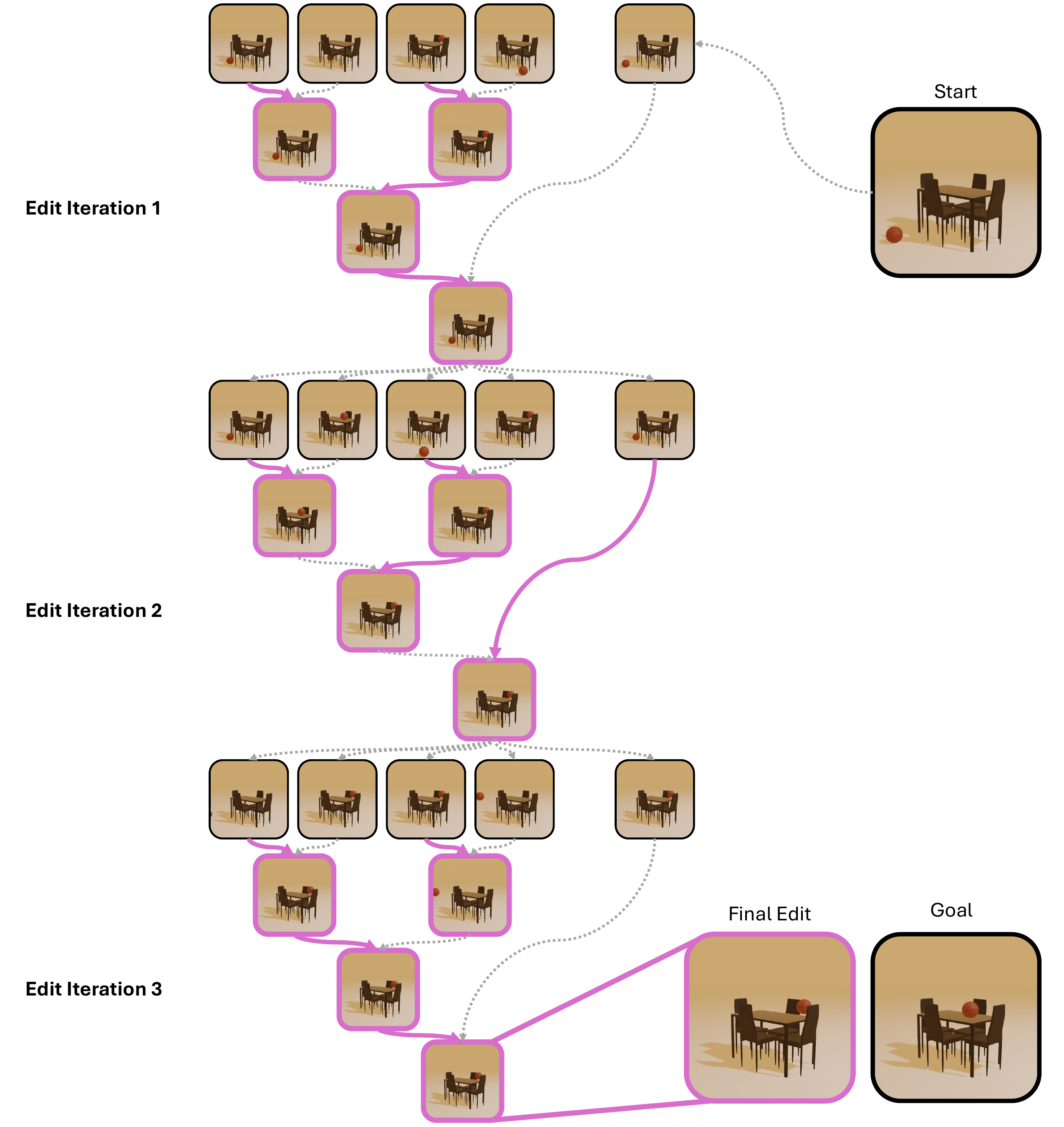}
\vspace{15pt}
\caption{A complete verification process of a 3x4 tree generated by GPT4o on one task instance. We observe that a more human-aligned candidate is generated in edit iteration 2 but is not selected by the verifier.}
\label{fig: veri_tree}
\end{figure}

\begin{figure}
\centering
\vspace{-75pt} 
\includegraphics[width=\linewidth, page=1]{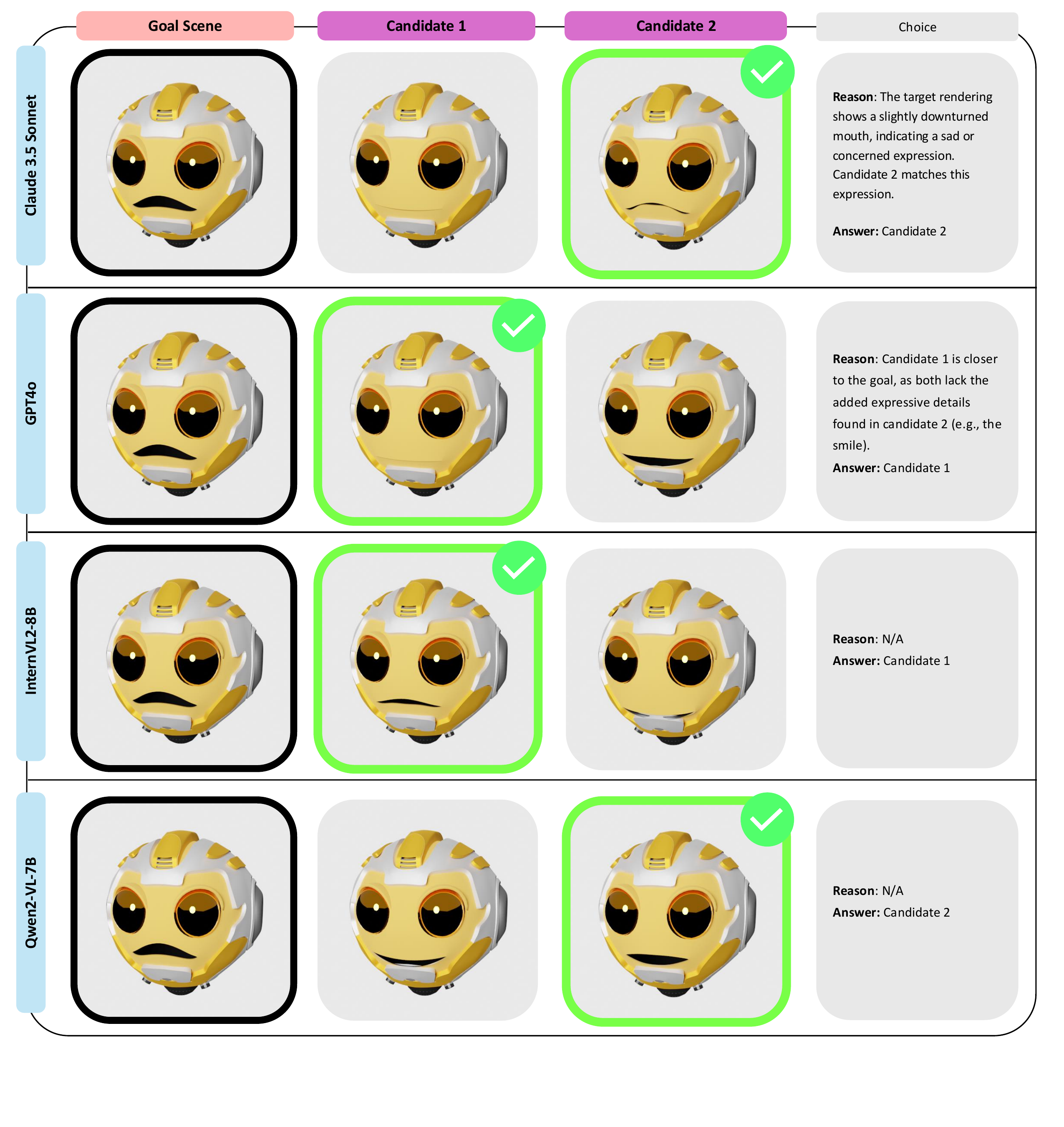}
\vspace{-45pt} 
\caption{Examples of verifier decisions for a blend shape instance. N/A indicates that no reasoning is provided by the verifier. The candidates differ across models since they are rendered from edits generated by the model itself.}
\label{fig: veri_fail_1}
\vspace{-10pt}
\end{figure}

\begin{figure}
\centering
\vspace{-75pt} 
\includegraphics[width=\linewidth, page=2]{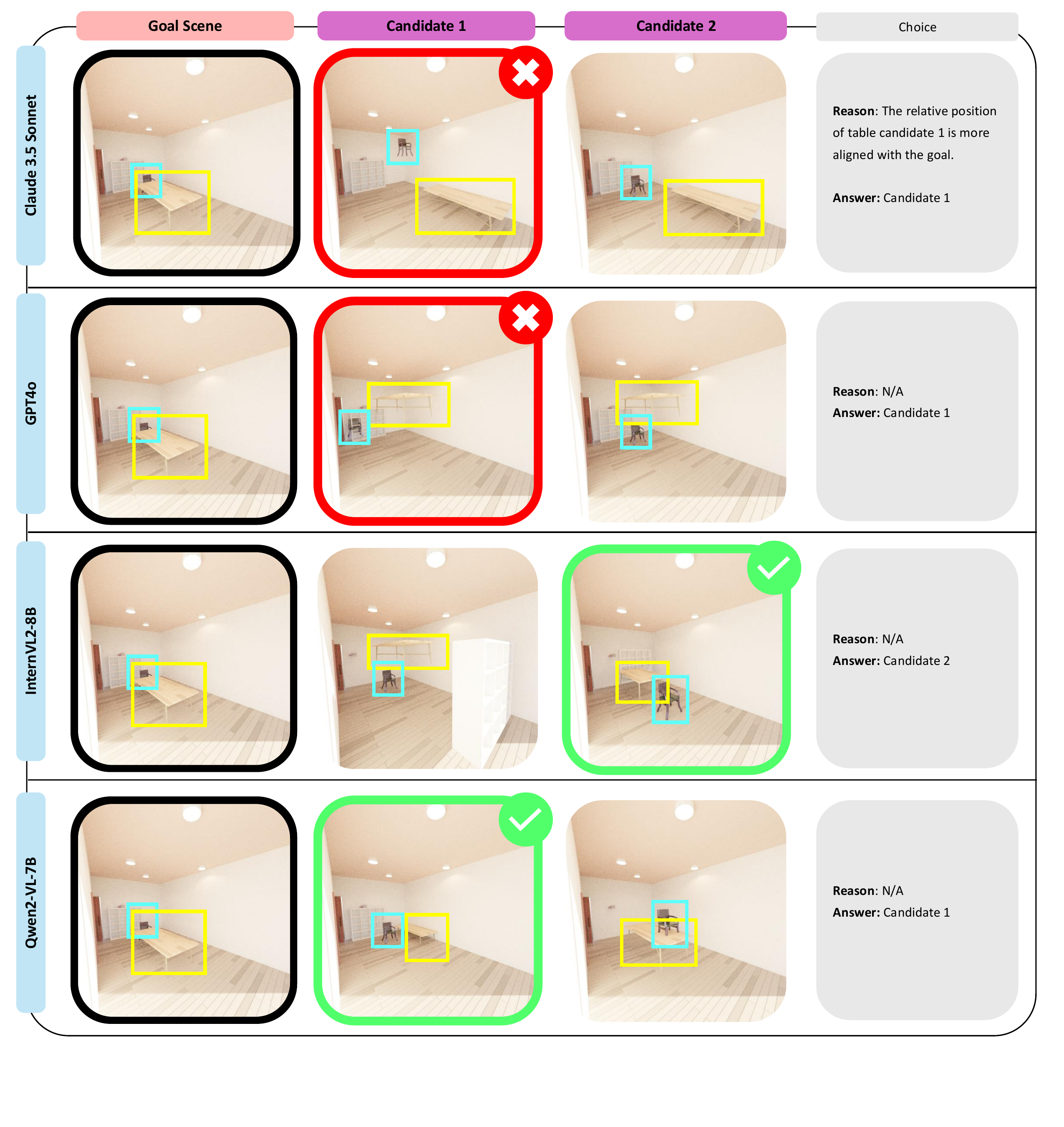}
\vspace{-45pt} 
\caption{Examples of verifier decisions for an object placement instance. N/A indicates that no reasoning is provided by the verifier. The candidates differ across models since they are rendered from edits generated by the model itself.}
\label{fig: veri_fail_2}
\vspace{-10pt}
\end{figure}

We provide a complete verification process of a 3x4 tree generated by GPT4o, shown in \cref{fig: veri_tree}, to contextualize the verification process. We also offer cross-model verifier comparisons on identical task instances in \cref{fig: veri_fail_1} and \cref{fig: veri_fail_2}. Despite the prompt guidance in \cref{fig: verifier}, only Claude 3.5 Sonnet consistently produces a complete reasoning process for its decisions, potentially enhancing its verification capability and contributing to its status as the most human-aligned VLM verifier.

\section{Calibration of Evaluation Metrics}
\label{sec: calibration}
\begin{figure}
\centering
\vspace{-15pt} 
\includegraphics[width=\linewidth, page=1]{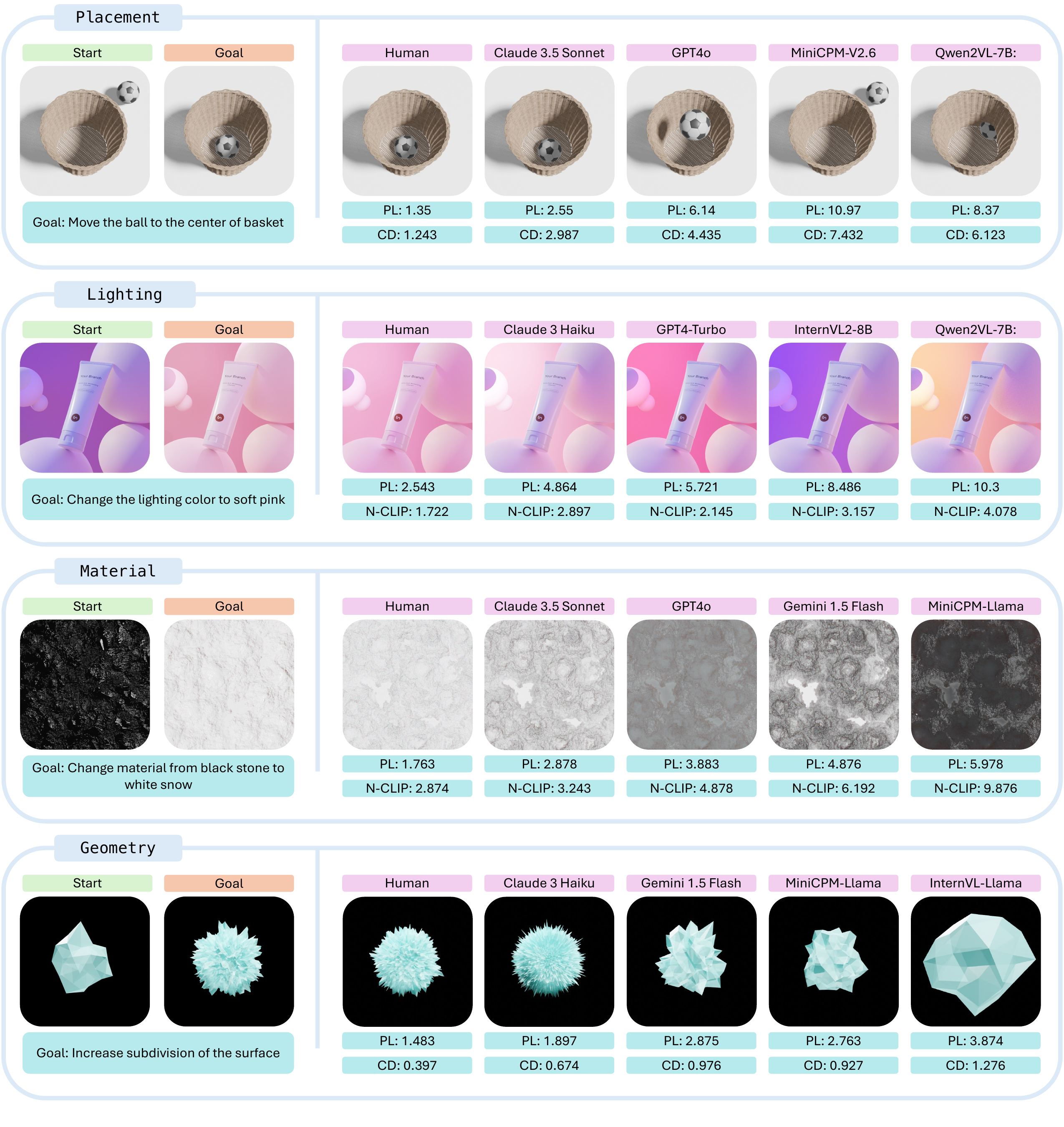}
\vspace{-13pt} 
\caption{Calibration of metric values with render images of VLM system output edits. We present start scene, goal scene, human user edit, and VLM system edits side by side with their corresponding metric values. }
\label{fig: calibration}
\end{figure}

We calibrate photometric loss (PL), negative-CLIP (N-CLIP), and Chamfer distance (CD) using the examples in \cref{fig: calibration} and Fig 3 of the main paper. PL and N-CLIP values are on the scale of $10^{-3}$ for blend shape and placement tasks and $10^{-2}$ for geometry, material, and lighting tasks. CD stays at its original scale. We notice that small metric differences can correspond to significant visual changes in the scene. 

While these metrics generally align with human perception, they have two key limitations: \vspace{-13pt} \paragraph{Failure in capturing physical plausibility} For instance, Qwen2VL-7B leaves the soccer ball unnaturally stuck on the basket, violating physical laws and common sense. In contrast, MiniCPM-V2.6 places the ball outside the basket. MiniCPM-V2.6's edit, despite being suboptimal, adheres to physical laws and should be considered superior to Qwen's edit, a distinction not captured by the metric scores.

\vspace{-13pt}\paragraph{Task-dependent disproportionate scale.} Lighting and procedural material tasks, due to their large-scale color changes, have higher values for N-CLIP and PL compared to blend shape and placement tasks. Procedural geometry editing also yields larger metric values since the object-of-interest often dominates the scene, making small changes more impactful. Conversely, placement and blend shape tasks typically involve object or feature adjustments of a smaller scale, leaving a significant proportion of the scene unchanged, leading to comparatively smaller metric values. Despite allowing cross-model comparison on a specific task, the disproportionate scales of metrics hinder direct cross-task comparison of a specific VLM system.

\section{Camera Viewpoint Selection}
\label{sec: camera view}
\begin{figure}[H]
\centering
\includegraphics[width=\linewidth, page=1]{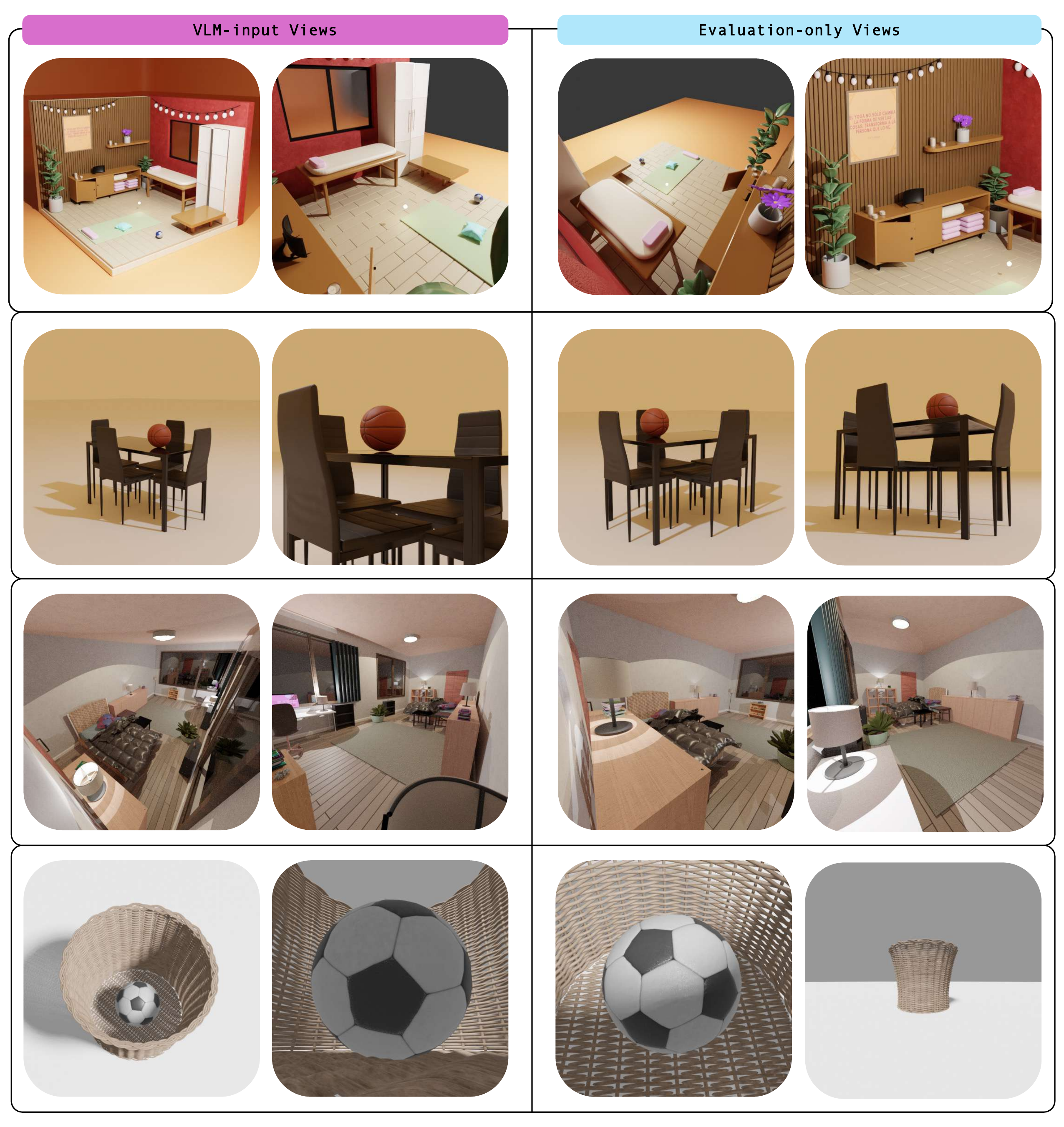}
\caption{Examples of VLM-input views and evaluation-only views. Images on the first column are all rendered from comprehensive views.}
\label{fig: multiview}
\end{figure}

\begin{figure}[H]
\centering
\includegraphics[width=\linewidth, page=2]{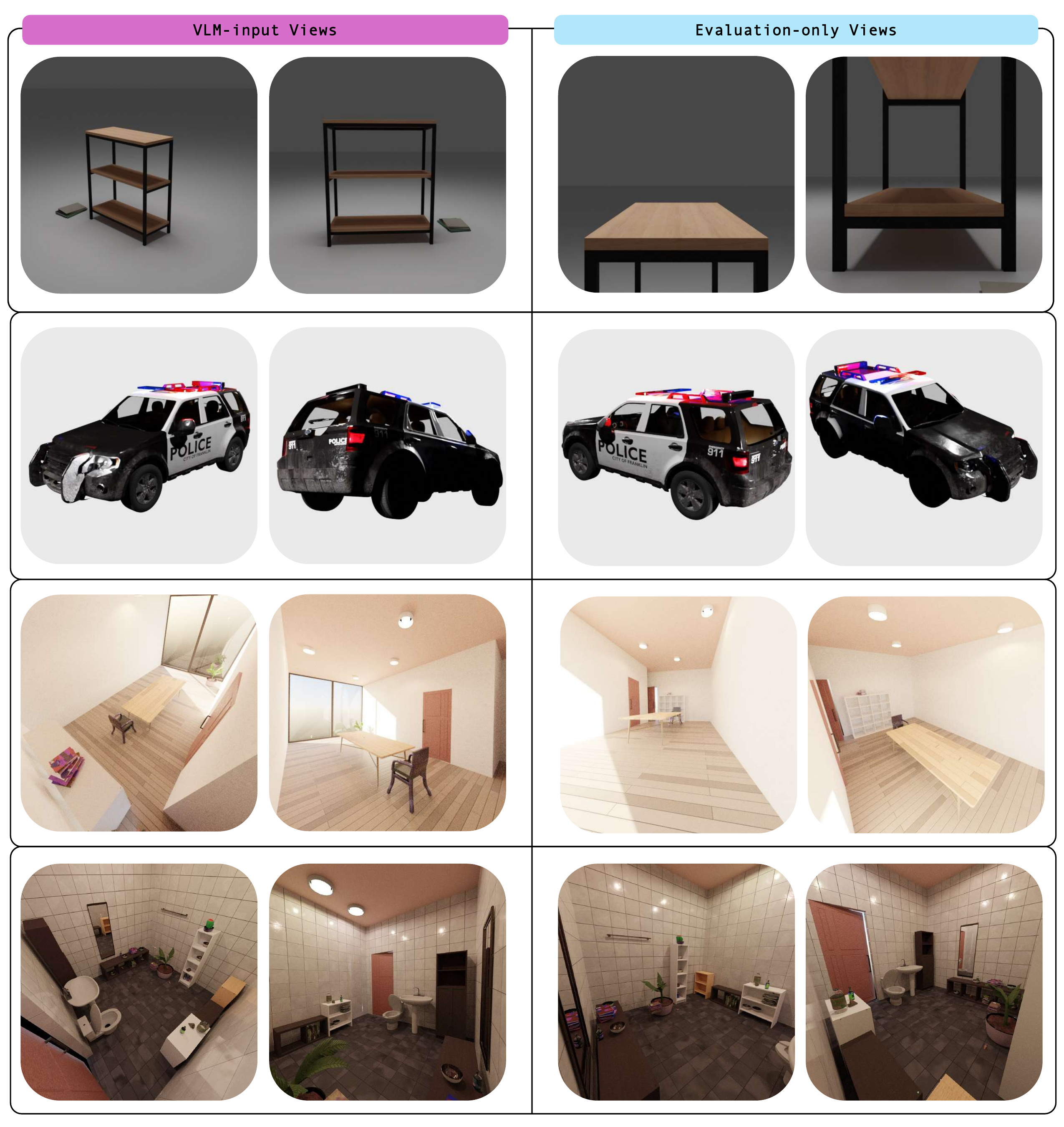}
\caption{Examples of VLM-input views and evaluation-only views. Images on the first column are all rendered from comprehensive views.}
\label{fig: multiview2}
\end{figure}

We define VLM-input and evaluation-only as two sets of views, with the former propagated to the VLM system and the latter reserved exclusively for evaluation. Both sets contribute to the evaluation metrics. A comprehensive view (defined below) is first selected and assigned to the VLM input set. Additional views capturing key object details are chosen, with one added to the VLM-input set and the remaining designated as evaluation-only. Importantly, all objects-of-interest are guaranteed to appear in at least one VLM-input view, ensuring the system has access to all critical visual information. Examples of this process are illustrated in \cref{fig: multiview} and \cref{fig: multiview2}.

We define a comprehensive view as a camera angle that provides a high-level perspective, typically from an elevated angle, encompassing most objects in the scene. It must clearly convey spatial relationships and object locations, particularly for objects-of-interest. While challenging to formalize in words, comprehensiveness is visually exemplified in \cref{fig: multiview} and \cref{fig: multiview2}, where the images in the first column are all comprehensive views.

\clearpage
\setlength{\intextsep}{0pt}
\setlength{\columnsep}{10pt}

\clearpage
\setlength{\intextsep}{0pt}
\setlength{\columnsep}{10pt}

\clearpage
\setlength{\intextsep}{0pt}
\setlength{\columnsep}{10pt}

\clearpage
\setlength{\intextsep}{0pt}
\setlength{\columnsep}{10pt}

\clearpage
\setlength{\intextsep}{0pt}
\setlength{\columnsep}{10pt}

\clearpage
\setlength{\intextsep}{0pt}
\setlength{\columnsep}{10pt}

\clearpage
\setlength{\intextsep}{0pt}
\setlength{\columnsep}{10pt}

\clearpage
\setlength{\intextsep}{0pt}
\setlength{\columnsep}{10pt}

\clearpage
\setlength{\intextsep}{0pt}
\setlength{\columnsep}{10pt}


\clearpage

\section{Prompts for VLM System}
\label{sec: prompts}
In our generator-verifier VLM system implementation, three VLM agents are involved: generator, code editor, and verifier. Here we include the prompt template we use for the three agents. 
\subsection{Brainstormer}
Brainstormer compares the start and goal render images, interprets the Python script of the start scene, and generates instructions for the required modifications on the script. It operates \textbf{alternatively} in two distinct modes: \textit{tune} and \textit{leap}, following from BlenderAlchemy. The \textit{tune} mode adjusts parameter values within the existing code, while the \textit{leap} mode proposes structural changes to the code, such as introducing new nodes for procedural editing. We set the return format to be a start-separated list of at most five instruction pieces. The prompt for both modes is given in \cref{fig: brainstormer_leap} and \cref{fig: brainstormer_tune}. 

\subsection{Code Editor}
The code editor iterates through the brainstormer's output list of instruction pieces and integrates each of them into the code script of start scene. It generates a list of Python code differences, each including ``CodeBefore,'' the original code segment from the input script, and ``CodeAfter,'' the corresponding proposed modification to be applied. We use some helper function subsequently substitute CodeBefore with CodeAfter.

\subsection{Verifier}
The verifier concatenates the render images of two proposal edits horizontally, compares them with the goal render, and selects one that is more similar to the goal edit. It returns a `left'' or ``right'' choice over the concatenated image,  indicating the choice among the two candidates.

\vspace{15pt}
\begin{figure*}[ht]
\centering
\begin{tcolorbox}[colback=gray!10, colframe=gray!80, width=\textwidth]

The following Blender code was used to produce a procedural 3D model: \\

\texttt{```} {\textbf{[Python script of the START scene]}} \texttt{```} \\

The final code creates a procedural 3D model and produces the rendering on the left below(The image is concatenated by camera renders from different angles):\\

The desired procedural 3D model is shown in the image on the right(The image is concatenated by camera renders from different angles). Please describe the difference between the two 3D models, and edit the code above to reflect this desired change. \\

\textbf{[A concatenated image of START(on the left) and GOAL(on the right)]}\\

Describe, in a bullet-point list (using * as the bullet points), the biggest visual difference, which lines you would change (quote them in python code blocks) and how you would change them. Every item of the list should reference only ONE or A FEW lines of code and how it should be changed. Make AT MOST 5 such changes, no more than 5. Return in the format below:\\

\indent raw: A new-line separated bullet point list that follows the following format:\\
    
     \indent Example: \\
     \indent * first item  \\
     \indent * second item  \\
     \indent ...etc  \\

\end{tcolorbox}
\caption{Prompt for brainstormer in \textit{leap} mode. This prompt is for procedural geometry editing task, but the ones for other tasks follow a similar structure with a few words changed.}
\label{fig: brainstormer_leap}
\end{figure*}

\begin{figure*}[ht]
\centering
\vspace{10pt}
\begin{tcolorbox}[colback=gray!10, colframe=gray!80, width=\textwidth]
The following Blender code was used to produce a procedural 3D model:\\

\texttt{```} {\textbf{[Python script of the START scene]}} \texttt{```} \\

This creates a procedural 3D model and produces the rendering on the left below (the image is concatenated by camera renders from different angles):\\

The desired 3D model is shown in the image on the right (the image is concatenated by camera renders from different angles). \\

\textbf{[A concatenated image of START(on the left) and GOAL(on the right)]}\\

Answer the following questions:\\
1) What is the SINGLE most visually obvious difference between the two models in the two renderings in the image above (both images are concatenated by camera renders from different angles)?\\
2) Look at the code. Which fields/variables which are set to numerical values are most likely responsible for the obvious visual difference in your answer to question 1?\\
3) Replace the assignments of such fields/variables accordingly!\\

Describe, in a bullet-point list (using * as the bullet points), the biggest visual difference, which lines you would change (quote them in python code blocks) and how you would change them. Every item of the list should reference only ONE or A FEW lines of code and how it should be changed. Make AT MOST 5 such changes, no more than 5. Return in the format below:\\

\indent gpt\_raw: A new-line separated bulletpoint list that follows the following format:\\
    
     \indent Example: \\
     \indent * first item  \\
     \indent * second item  \\
     \indent ...etc  \\

\end{tcolorbox}
\caption{Prompt for brainstormer in \textit{tune} mode. This prompt is for procedural geometry editing task, but the ones for other tasks follow a similar structure with a few words changed.}
\label{fig: brainstormer_tune}
\end{figure*}

\clearpage

\begin{figure*}[ht]
\centering
\begin{tcolorbox}[colback=gray!10, colframe=gray!80, width=\textwidth]

Consider the following code of a procedural 3D model in Blender:\\

\texttt{```} {\textbf{[Python script of the START scene]}} \texttt{```} \\

You'd like to do the following:\\

\textbf{[Instruction piece from brainstormer]
}\\

Convert this into a concrete code difference indicated by ``Before:" and ``After:" labels,
followed by code blocks that indicate which line should be
changed and to what. Do not copy-paste the whole original code.\\

Example:\\

Before:\\
\texttt{```} \\
python\\
a = 1\\
\texttt{```} \\
After: \\
\texttt{```} \\
python \\
a = 2  \\
\texttt{```} \\

\end{tcolorbox}
\vspace{-8pt}
\caption{Prompt for code editor. It receives instruction from brainstormer and incorporates it to the code script of start scene.}
\label{fig: code_editor}
\end{figure*}

\begin{figure*}[ht]
\vspace{30pt}

\centering
\begin{tcolorbox}[colback=gray!10, colframe=gray!80, width=\textwidth]

Here is the goal model rendering (the image is concatenated by camera renders from different angles):\\

\textbf{[goal\_model\_image]}\\

Below, I show two different models (the images are concatenated by camera renders from different angles). Which one is visually more similar to the goal model rendering?\\

\textbf{[A concatenated image of Candidate 1 (on the left) and Candidate 2 (on the right)]}\\

Return your answer in the following format: \\

raw: A block of text that contains a single word in a text block, indicated by \texttt{```} \text{.} The word should be either ``right" or ``left". Example:\\

You've asks me to choose the image (left or right) that best aligns with the goal render.\\
Though the sample on the left is more realistic, the sample on the right is better aligned with the goal render.\\
\texttt{```} \\
right\\
\texttt{```} \\

\end{tcolorbox}
\vspace{-8pt}
\caption{Prompt for verifier.}
\label{fig: verifier}
\end{figure*}



{
    \small
    \bibliographystyle{ieeenat_fullname}
    \bibliography{main}
}



\end{document}